\documentclass[10pt]{IEEEtran}
\usepackage[LGR,T1]{fontenc}
\usepackage[latin9]{inputenc}
\usepackage{color}
\usepackage{amsmath}
\usepackage{amsthm}
\usepackage{amssymb}

\makeatletter
\theoremstyle{plain}
\newtheorem{thm}{\protect\theoremname}
\theoremstyle{remark}
\newtheorem{rem}[]{\protect\remarkname}
\theoremstyle{plain}
\newtheorem{lem}[]{\protect\lemmaname}
\theoremstyle{plain}
\newtheorem{fact}[]{\protect\factname}

\usepackage{cite}

\usepackage[mathscr]{eucal}
\usepackage{fixltx2e}
\usepackage{array}
\usepackage{verbatim}
\usepackage{bm}
\usepackage{algorithmic}
\usepackage{algorithm}
\usepackage{verbatim}
\usepackage{textcomp}
\usepackage{mathrsfs}
\usepackage{epstopdf}

\newcommand{\openone}{\leavevmode\hbox{\small1\normalsize\kern-.33em1}}

\catcode`~=11 \def\UrlSpecials{\do\~{\kern -.15em\lower .7ex\hbox{~}\kern .04em}} \catcode`~=13

\allowdisplaybreaks[4]

\newcommand{\calA}{\mathcal{A}}
\newcommand{\calB}{\mathcal{B}}

\newcommand{\calK}{\mathcal{K}}

\newcommand{\calM}{\mathcal{M}}

\newcommand{\calP}{\mathcal{P}}
\newcommand{\calQ}{\mathcal{Q}}

\newcommand{\calT}{\mathcal{T}}
\newcommand{\calU}{\mathcal{U}}
\newcommand{\calV}{\mathcal{V}}

\newcommand{\calX}{\mathcal{X}}
\newcommand{\calY}{\mathcal{Y}}

\newcommand{\rme}{\mathrm{e}}

\newcommand{\bbR}{\mathbb{R}}

\DeclareMathAlphabet{\mathbsf}{OT1}{cmss}{bx}{n}
\DeclareMathAlphabet{\mathssf}{OT1}{cmss}{m}{sl}

\DeclareSymbolFont{bsfletters}{OT1}{cmss}{bx}{n}
\DeclareSymbolFont{ssfletters}{OT1}{cmss}{m}{n}
\DeclareMathSymbol{\bsfGamma}{0}{bsfletters}{'000}
\DeclareMathSymbol{\ssfGamma}{0}{ssfletters}{'000}
\DeclareMathSymbol{\bsfDelta}{0}{bsfletters}{'001}
\DeclareMathSymbol{\ssfDelta}{0}{ssfletters}{'001}
\DeclareMathSymbol{\bsfTheta}{0}{bsfletters}{'002}
\DeclareMathSymbol{\ssfTheta}{0}{ssfletters}{'002}
\DeclareMathSymbol{\bsfLambda}{0}{bsfletters}{'003}
\DeclareMathSymbol{\ssfLambda}{0}{ssfletters}{'003}
\DeclareMathSymbol{\bsfXi}{0}{bsfletters}{'004}
\DeclareMathSymbol{\ssfXi}{0}{ssfletters}{'004}
\DeclareMathSymbol{\bsfPi}{0}{bsfletters}{'005}
\DeclareMathSymbol{\ssfPi}{0}{ssfletters}{'005}
\DeclareMathSymbol{\bsfSigma}{0}{bsfletters}{'006}
\DeclareMathSymbol{\ssfSigma}{0}{ssfletters}{'006}
\DeclareMathSymbol{\bsfUpsilon}{0}{bsfletters}{'007}
\DeclareMathSymbol{\ssfUpsilon}{0}{ssfletters}{'007}
\DeclareMathSymbol{\bsfPhi}{0}{bsfletters}{'010}
\DeclareMathSymbol{\ssfPhi}{0}{ssfletters}{'010}
\DeclareMathSymbol{\bsfPsi}{0}{bsfletters}{'011}
\DeclareMathSymbol{\ssfPsi}{0}{ssfletters}{'011}
\DeclareMathSymbol{\bsfOmega}{0}{bsfletters}{'012}
\DeclareMathSymbol{\ssfOmega}{0}{ssfletters}{'012}

\newcommand{\tilC}{\tilde{C}}

\newcommand{\eps}{\varepsilon}

\usepackage{graphicx}

\def\e{{\rm e}}

\allowdisplaybreaks[1]
\providecommand{\factname}{Fact}
\providecommand{\lemmaname}{Lemma}
\providecommand{\remarkname}{Remark}
\providecommand{\theoremname}{Theorem}

\newcounter{mytempeqncnt}

\makeatother

\begin{document}

\title{Wyner's Common Information under R\'enyi Divergence Measures}

\author{Lei Yu and Vincent Y. F. Tan, \IEEEmembership{Senior Member,~IEEE}
\thanks{Manuscript received September 11, 2017; revised December 07, 2017,
January 25, 2018; accepted January 25, 2018.  This work was supported
by the Singapore National Research Foundation (NRF) National Cybersecurity
R\&D Grant under Grants R-263-000-C74-281 and NRF2015NCR-NCR003-006.}
\thanks{
L.~Yu is  with the Department of Electrical and Computer Engineering,
National University of Singapore (NUS), Singapore 117583 (e-mail: leiyu@nus.edu.sg).
V.~Y.~F.~Tan is  with the with the Department of Electrical and Computer Engineering and the Department of Mathematics, NUS, Singapore 119076 (e-mail: vtan@nus.edu.sg).}
\thanks{
Communicated by A. Khisti, Associate Editor for Shannon Theory. }
\thanks{Copyright
(c) 2018 IEEE. Personal use of this material is permitted. However,
permission to use this material for any other purposes must be obtained
from the IEEE by sending a request to pubs-permissions@ieee.org.}}
\maketitle
\begin{abstract}
We study a generalized version of Wyner's common information problem
(also coined the distributed source simulation problem). The original
common information problem consists in understanding the minimum rate
of the common input to independent processors to generate an approximation
of a joint distribution when the distance measure used to quantify
the discrepancy between the synthesized and target distributions is
the normalized relative entropy. Our generalization involves changing
the distance measure to the unnormalized and normalized R\'enyi divergences
of order $\alpha=1+s\in[0,2]$. We show that the minimum rate needed
to ensure the R\'enyi divergences between the distribution induced by
a code and the target distribution vanishes remains the same as the
one in Wyner's setting, except when the order $\alpha=1+s=0$. This
implies that Wyner's common information is rather robust to the choice
of distance measure employed. As a byproduct of the proofs used to
the establish the above results, the exponential strong converse for
the common information problem under the total variation distance
measure is established.
\end{abstract}


\section{\label{sec:Introduction}Introduction}

How much common randomness is needed to simulate two correlated sources
in a distributed fashion? This problem, termed {\em distributed
source simulation}, was first studied by Wyner \cite{Wyner}, who
used the normalized relative entropy (Kullback-Leibler divergence
or KL divergence) to measure the approximation level (discrepancy)
between the simulated joint distribution and the joint distribution
of the original correlated sources. He defined the minimum rate needed
to ensure that the normalized relative entropy vanishes asymptotically
as the \emph{common information} between the sources. He also established
a single-letter characterization for the common information, i.e.,
the common information between correlated sources $X$ and $Y$ (with
target distribution $\pi_{XY}$) is
\begin{equation}
C_{\mathsf{Wyner}}(X;Y)=\min_{P_{XYW}:\,P_{XY}=\pi_{XY},\,X-W-Y}I(XY;W).\label{eqn:CWyner}
\end{equation}
The common information is also known to be one of many reasonable
measures of the dependence between two random variables~\cite[Section~14.2.2]{Gamal}
(other measures include the mutual information and the G\'acs-K\"orner-Witsenhausen
common information). A related notion is that of the \emph{exact common
information} which was introduced by Kumar, Li, and El Gamal~\cite{Kumar}.
They assumed variable-length codes and exact generation of the correlated
sources $(X,Y)$, instead of block codes and approximate simulation
of $\pi_{XY}$ as assumed by Wyner \cite{Wyner}. The exact common
information is not smaller than Wyner's common information. However,
it is still not known whether they are equal in general. Furthermore,
the common information problem can be also be regarded as a distributed
coordination problem. The concept of \emph{coordination} was first
introduced by Cuff, Permuter, and Cover \cite{Cuff10,Cuff}, who used
the total variation (TV) distance to measure the level of approximation
between the simulated and target distributions.

Wyner's common information problem is also closely related to the
channel resolvability problem, which was first studied by Han and
Verd\'u \cite{Han}, and subsequently studied by Hayashi \cite{Hayashi06,Hayashi11},
Liu, Cuff, and Verd\'u \cite{Liu}, and Yu and Tan \cite{Yu} among
others. For the achievability part, both problems rely on so-called
soft-covering lemmas~\cite{Cuff}. The channel resolvability or common
information problems have several interesting applications\textemdash including
secrecy, channel synthesis, and source coding. For example, in \cite{Bloch}
it was used to study the performance of a wiretap channel system under
different secrecy measures. In~\cite{Han14} it was used to study
the reliability and secrecy exponents of a wiretap channel with cost
constraints. In \cite{Parizi} it was used to study the exact secrecy
and reliability exponents for a wiretap channel.

\subsection{Main Contributions}

Different from Wyner's work, we use (normalized and unnormalized)
R\'enyi divergences of order $1+s\in[0,2]$ to measure the level of
approximation between the simulated and target distributions. This
is motivated in part by our desire to understand the sensitivity of
the divergence as approximation measure on Wyner's common information.
We prove that for the distributed source simulation problem, the minimum
rate needed to guarantee that the (normalized and unnormalized) R\'enyi
divergences vanish asymptotically is equal to Wyner's common information
(except for the case when R\'enyi parameter is equal to~$0$). This
implies that Wyner's common information in \eqref{eqn:CWyner} is
rather robust to the distance measure. For the achievability part,
by using the method of types and typicality arguments, we prove that
the optimal R\'enyi divergences vanish (at least) exponentially fast
if the code rate is larger than Wyner's common information. However,
for the converse part, the proof is not straightforward and we have
to first consider an auxiliary problem. We first prove an {\em exponential
strong converse} for the common information problem under the TV
distance measure, i.e., when the code rate is smaller than Wyner's
common information, the TV distance between the induced distribution
and the target distribution tends to one (at least) exponentially
fast. Even though our proof technique mirrors that of Oohama~\cite{oohama2016exponent}
to establish the exponential strong converse for the Wyner-Ziv problem,
it differs significantly in some aspects. To wit, some intricate continuity
arguments are required to assert that the strong converse exponent
is positive for all rates below $C_{\mathsf{Wyner}}(X;Y)$ (see part
(i) of Lemma \ref{propFOmega}). Furthermore and interestingly, by
leveraging a key relationship between the R\'enyi divergence and the
TV distance \cite{sason2016renyi}, this exponential strong converse
implies the converse for the normalized R\'enyi divergence (which in
turn also implies the strong converse for the unnormalized R\'enyi divergence).

It is worth noting that it is quite natural to use various divergences
to measure the discrepancy between two distributions. Wyner \cite{Wyner}
used the KL divergence to measure the level of approximation in the
distributed source synthesis problem; Hayashi~\cite{Hayashi06,Hayashi11}
and Yu and Tan \cite{Yu} respectively used the KL divergence and
the R\'enyi divergence to study the channel resolvability problem. The
latter also applied their results to study the capacity region for
the wiretap channel under these generalized measures. Furthermore,
in probability theory, Barron \cite{barron1986entropy} and Bobkov,
Chistyakov and G\"otze \cite{bobkov2016r} respectively used the KL
divergence and the R\'enyi divergence to study the central limit theorem,
i.e., they used them to measure the discrepancy between the induced
distribution of sum of i.i.d.\ random variables and the normal distribution
with the same mean and variance. Furthermore, special instances of
R\'enyi entropies and divergences\textemdash including the KL divergence,
the collision entropy (the R\'enyi divergence of order $2$), and min-entropy
(the R\'enyi divergence of order $\infty$)\textemdash were used to
study various information-theoretic problems (including security,
cryptography, and quantum information) in several works in the recent
literature~\cite{Bloch,hou2014effective,Yu,beigi2014quantum,dodis2013overcoming,Hayashi17,Tan}.

\subsection{Notation }

We use $P_{X}(x)$ to denote the probability distribution of a random
variable $X$. This will also be denoted as $P(x)$ (when the random
variable $X$ is clear from the context). We also use $\widetilde{P}_{X}$,
$\widehat{P}_{X}$ and $Q_{X}$ to denote various probability distributions
with alphabet $\mathcal{X}$. All alphabets considered in the sequel
are finite. The set of probability measures on $\mathcal{X}$ is denoted
as $\mathcal{P}\left(\mathcal{X}\right)$, and the set of conditional
probability measures on $\mathcal{Y}$ given a variable in $\mathcal{X}$
is denoted as $\mathcal{P}(\mathcal{Y}|\mathcal{X}):=\left\{ P_{Y|X}:P_{Y|X}(\cdot|x)\in\mathcal{P}(\mathcal{Y}),x\in\mathcal{X}\right\} $.
Furthermore, the support of a distribution $P\in\calP(\calX)$ is
denoted as $\mathrm{supp}(P)=\{x\in\calX:P(x)>0\}$.

We use $T_{x^{n}}(x):=\frac{1}{n}\sum_{i=1}^{n}1\left\{ x_{i}=x\right\} $
to denote the type (empirical distribution) of a sequence $x^{n}$,
$T_{X}$ and $V_{Y|X}$ to respectively denote a type of sequences
in $\mathcal{X}^{n}$ and a conditional type of sequences in $\mathcal{Y}^{n}$
(given a sequence $x^{n}\in\calX^{n}$). For a type $T_{X}$, the
type class (set of sequences having the same type $T_{X}$) is denoted
by $\mathcal{T}_{T_{X}}$. For a conditional type $V_{Y|X}$ and a
sequence $x^{n}$, the $V_{Y|X}$\emph{-shell of $x^{n}$} (the set
of $y^{n}$ sequences having the same conditional type $V_{Y|X}$
given $x^{n}$) is denoted by $\mathcal{T}_{V_{Y|X}}(x^{n})$. For
brevity, sometimes we use $T(x,y)$ to denote the joint distributions
$T(x)V(y|x)$ or $T(y)V(x|y)$.

The $\epsilon$-typical set of $Q_{X}$ is denoted as
\begin{align}
 \mathcal{T}_{\epsilon}^{n}(Q_{X})&:=\big\{ x^{n}\in\mathcal{X}^{n} :\nonumber \\
 & \left|T_{x^{n}}(x)-Q_{X}(x)\right|\leq\epsilon Q_{X}(x),\forall x\in\mathcal{X}\big\} .\label{eqn:typ_set}
\end{align}
The conditionally $\epsilon$-typical set of $Q_{XY}$ is denoted
as
\begin{equation}
\mathcal{T}_{\epsilon}^{n}(Q_{YX}|x^{n}):=\left\{ y^{n}\in\mathcal{Y}^{n}:(x^{n},y^{n})\in\mathcal{T}_{\epsilon}^{n}(Q_{XY})\right\} .
\end{equation}
For brevity, sometimes we write $\mathcal{T}_{\epsilon}^{n}(Q_{X})$
and $\mathcal{T}_{\epsilon}^{n}(Q_{YX}|x^{n})$ as $\mathcal{T}_{\epsilon}^{n}$
and $\mathcal{T}_{\epsilon}^{n}(x^{n})$ respectively.

The TV distance between two probability mass functions $P$ and $Q$
with a common alphabet $\calX$ is defined as
\begin{equation}
|P-Q|:=\frac{1}{2}\sum_{x\in\calX}|P(x)-Q(x)|.
\end{equation}
By the definition of $\epsilon$-typical set, we have that for any
$x^{n}\in\mathcal{T}_{\epsilon}^{n}(Q_{X})$,
\begin{equation}
\left|T_{x^{n}}-Q_{X}\right|\leq\frac{\epsilon}{2}.
\end{equation}

Fix distributions $P_{X},Q_{X}\in\calP(\calX)$. The {\em relative
entropy} and the {\em R\'enyi divergence of order $1+s$} are respectively
defined as
\begin{align}
D(P_{X}\|Q_{X}) & :=\sum_{x\in\mathrm{supp}(P_{X})}P_{X}(x)\log\frac{P_{X}(x)}{Q_{X}(x)}\label{eq:-19}\\*
D_{1+s}(P_{X}\|Q_{X}) & :=\frac{1}{s}\log\sum_{x\in\mathrm{supp}(P_{X})}P_{X}(x)^{1+s}Q_{X}(x)^{-s},\label{eq:-40}
\end{align}
and the conditional versions are respectively defined as
\begin{align}
D(P_{Y|X}\|Q_{Y|X}|P_{X}) & :=D(P_{X}P_{Y|X}\|P_{X}Q_{Y|X})\\*
D_{1+s}(P_{Y|X}\|Q_{Y|X}|P_{X}) & :=D_{1+s}(P_{X}P_{Y|X}\|P_{X}Q_{Y|X}),
\end{align}
where the summations in \eqref{eq:-19} and \eqref{eq:-40} are taken
over the elements in $\mathrm{supp}(P_{X})$. Throughout, $\log$
is to the natural base $\rme$ and $s\geq-1$. It is known that $\lim_{s\to0}D_{1+s}(P_{X}\|Q_{X})=D(P_{X}\|Q_{X})$
so a special case of the R\'enyi divergence (or the conditional version)
is the usual relative entropy (or the conditional version).

Given a number $a\in[0,1]$, we define $\bar{a}=1-a$. We also define
$\left[x\right]^{+}=\max\left\{ x,0\right\} $.

\subsection{Problem Formulation}

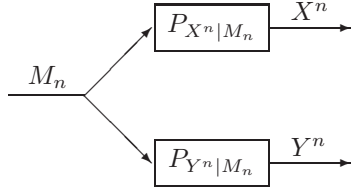
\begin{figure}
\centering \setlength{\unitlength}{0.05cm} 
{ \begin{picture}(100,60) 
\put(5,30){\line(1,0){20}} \put(25,30){\vector(1,1){18}}
\put(25,30){\vector(1,-1){18}} \put(44,42){\framebox(30,12){$P_{X^{n}|M_{n}}$}}
\put(44,6){\framebox(30,12){$P_{Y^{n}|M_{n}}$}} \put(74,48){\vector(1,0){22}}
\put(74,12){\vector(1,0){22}} \put(10,33){%
\mbox{%
$M_{n}$%
}} \put(80,50){%
\mbox{%
$X^{n}$%
}} \put(80,14){%
\mbox{%
$Y^{n}$%
}} \end{picture}} \caption{Distributed source synthesis problem, where the random variable $M_{n}\in\calM_{n}:=\{1,\ldots,\e^{nR}\}$. }
\label{fig:dss}
\end{figure}

In this paper, we consider the distributed source simulation problem
illustrated in Fig.~\ref{fig:dss}. Given a target distribution $\pi_{XY}$,
we wish to minimize the alphabet size of a random variable $M_{n}$
that is uniformly distributed over\footnote{For simplicity, we assume that $\e^{nR}$ and similar expressions
are integers.} $\calM_{n}:=\{1,\ldots,\e^{nR}\}$ ($R$ is a positive number known
as the {\em rate}), such that the generated (or synthesized) distribution
\begin{align}
 & P_{X^{n}Y^{n}}(x^{n},y^{n})\nonumber \\
 & \qquad:=\frac{1}{|{\cal M}_{n}|}\sum_{m\in{\cal M}_{n}}P_{X^{n}|M_{n}}(x^{n}|m)P_{Y^{n}|M_{n}}(y^{n}|m)\label{eq:-113}
\end{align}
forms a good approximation to the product distribution $\pi_{X^{n}Y^{n}}:=\pi_{XY}^{n}$.
The pair of random mappings $(P_{X^{n}|M_{n}},P_{Y^{n}|M_{n}})$ constitutes
a \emph{synthesis code}.

Different from Wyner's seminal work on the distributed source simulation
problem~\cite{Wyner}, we employ the unnormalized R\'enyi divergence
\begin{equation}
D_{1+s}(P_{X^{n}Y^{n}}\|\pi_{X^{n}Y^{n}})\label{eq:-11}
\end{equation}
and the normalized R\'enyi divergence
\begin{equation}
\frac{1}{n}D_{1+s}(P_{X^{n}Y^{n}}\|\pi_{X^{n}Y^{n}})\label{eq:-11-1}
\end{equation}
to measure the discrepancy between $P_{X^{n}Y^{n}}$ and $\pi_{X^{n}Y^{n}}$.
The minimum rates required to ensure these two measures vanish asymptotically
are respectively termed the\emph{ unnormalized and normalized R\'enyi
common information}, and denoted as
\begin{align}
 & T_{1+s}(\pi_{XY})\nonumber \\
 & \quad:=\inf\left\{ R:\;\lim_{n\to\infty}D_{1+s}(P_{X^{n}Y^{n}}\|\pi_{X^{n}Y^{n}})=0\right\} ,\\
 & \widetilde{T}_{1+s}(\pi_{XY})\nonumber \\
 & \quad:=\inf\Big\{ R:\;\lim_{n\to\infty}\frac{1}{n}D_{1+s}(P_{X^{n}Y^{n}}\|\pi_{X^{n}Y^{n}})=0\Big\}.
\end{align}
It is clear that
\begin{equation}
\widetilde{T}_{1+s}(\pi_{XY})\le{T}_{1+s}(\pi_{XY}).\label{eqn:stronger}
\end{equation}
We also denote the minimum rate required to ensure the TV distance
is bounded above by some constant $\eps\in[0,1]$ asymptotically as
\begin{align}
 & T_{\eps}^{\mathsf{TV}}(\pi_{XY})\nonumber \\
 & \quad:=\inf\Big\{ R:\;\limsup_{n\to\infty}|P_{X^{n}Y^{n}}-\pi_{X^{n}Y^{n}}|\leq\eps\Big\}.
\end{align}
We say that the {\em strong converse property} for the common information
problem under the TV distance holds if $T_{\eps}^{\mathsf{TV}}(\pi_{XY})$
does not depend on $\eps\in[0,1)$.

\section{Main Results}

Our main result concerns Wyner's common information problem when the
discrepancy measure is the unnormalized or normalized R\'enyi divergence.
It is stated as follows.
\begin{thm}[R\'enyi Common Informations]
\label{thm:RenyiCI} The unnormalized and normalized and R\'enyi common
informations satisfy
\begin{align}
T_{1+s}(\pi_{XY}) & =\widetilde{T}_{1+s}(\pi_{XY})\\
 & =\begin{cases}
C_{\mathsf{Wyner}}(X;Y) & s\in(-1,1]\\
0 & s=-1
\end{cases}.
\end{align}
Furthermore, for $s\in(-1,1]$, the optimal R\'enyi divergence $D_{1+s}(P_{X^{n}Y^{n}}\|\pi_{X^{n}Y^{n}})$
in the definitions of the R\'enyi common informations decays at least
exponentially fast in $n$ when $R>C_{\mathsf{Wyner}}(X;Y)$.
\end{thm}
\begin{rem}
For the converse part, $T_{1+s}(\pi_{XY})\geq\widetilde{T}_{1+s}(\pi_{XY})\geq C_{\mathsf{Wyner}}(X;Y)$
for $s\in[0,1]$ is implied by Wyner's work \cite{Wyner} and the
monotonicity of the R\'enyi divergence. For the achievability part,
$\widetilde{T}_{1+s}(\pi_{XY})\leq C_{\mathsf{Wyner}}(X;Y)$ for $s\in(-1,0]$
is also implied by Wyner's work \cite{Wyner} and the monotonicity
of the R\'enyi divergence. Furthermore, since a channel resolvability
code for the memoryless channel $P_{X|W}\times P_{Y|W}$ can be used
to form a common information code, the achievability part for the
common information problem can be obtained from existing channel resolvability
results. Specifically, $T_{1+s}(\pi_{XY})\leq C_{\mathsf{Wyner}}(X;Y)$
for $s\in(-1,0]$ can be obtained from Hayashi's \cite{Hayashi06,Hayashi11}
or Han, Endo, and Sasaki's results \cite{Han14}. In addition, $\widetilde{T}_{1+s}(\pi_{XY})\leq T_{1+s}(\pi_{XY})\leq C_{1+s}\left(X;Y\right)$
for $s\in(0,1]$ with
\begin{align}
 & C_{1+s}\left(X;Y\right):=\min_{P_{XYW}:\,P_{XY}=\pi_{XY},\,X-W-Y}\nonumber \\
 & \qquad\sum_{w}P_{W}\left(w\right)D_{1+s}\left(P_{X|W}(\cdot|w)P_{Y|W}(\cdot|w)\|P_{XY}\right)
\end{align}
can be obtained from the present authors' results \cite{Yu}, but
as shown in Theorem \ref{thm:RenyiCI}, this bound is not tight since
$C_{1+s}\left(X;Y\right)>C_{\mathsf{Wyner}}(X;Y)$ in general for
$s\in(0,1]$. This is because, on the one hand, for the channel resolvability
problem, the discrete memoryless channel is fixed, and, by construction,
imposes a product conditional distribution of the output given the
input (which is a product distribution), but for the common information
problem, the synthesizer has the freedom to choose $P_{X^{n}Y^{n}|M_{n}}=P_{X^{n}|M_{n}}\times P_{Y^{n}|M_{n}}$,
so that the Markov chain $X^{n}-M_{n}-Y^{n}$ holds; on the other
hand, for the common information problem, in the sequel, we will show
that if we utilize a truncated channel (which is not memoryless) as
the synthesizer. This results in a smaller achievable rate for the
case $s\in(0,1]$. Therefore, our converse for $s\in[-1,0)$ and achievability
for $s\in(0,1]$ are new (and also tight).
\end{rem}
\begin{rem}
An exponential achievability result for $s\in(-1,0]$ can be obtained
from Hayashi's \cite{Hayashi06,Hayashi11} and Han, Endo, and Sasaki's
results~\cite{Han14}, where i.i.d.\ codes were employed.
\end{rem}
For this theorem, the proof of the achievability part for the unnormalized
R\'enyi common information is provided in Appendix \ref{sec:Proof-of-Theoremunnormalized},
and the proof of the converse part for the normalized R\'enyi common
information is provided in Section \ref{sec:conversefornormRenyi}.
Observe that the unnormalized R\'enyi divergence is stronger than the
normalized one in the sense of \eqref{eqn:stronger}, hence $\widetilde{T}_{1+s}(\pi_{XY})\leq{T}_{1+s}(\pi_{XY})$.
This implies, on one hand, the achievability result for the normalized
R\'enyi common information $\widetilde{T}_{1+s}(\pi_{XY})$ can be obtained
directly from the achievability result for the unnormalized version
${T}_{1+s}(\pi_{XY})$, and on the other hand, the converse result
for the normalized R\'enyi common information $\widetilde{T}_{1+s}(\pi_{XY})$
implies the converse result for the unnormalized version ${T}_{1+s}(\pi_{XY})$.

The R\'enyi common informations are the same for all $s\in(-1,1]$,
and also same as Wyner's common information $C_{\mathsf{Wyner}}(X;Y)$
(which corresponds to $s=0$ for the normalized case). For the case
$s\in(-1,1]$, to obtain the (unnormalized and normalized) R\'enyi common
informations, we utilize a random code with $(W^{n},X^{n},Y^{n})$
($W$ is the auxiliary random variable in the definition of $C_{\mathsf{Wyner}}(X;Y)$)
distributed according to a truncated product distribution, i.e., a
product distribution governed by $Q_{WXY}^{n}$ but whose mass is
truncated to the typical set $\mathcal{T}_{\epsilon}^{n}(Q_{WXY})$.\footnote{Interestingly, a truncated code is not necessary for $s\in(-1,0]$
case, since an i.i.d.\ code (without truncation) is optimal as well
for this case. This point can be seen from the work of Yu and Tan~\cite{Yu}.} On one hand, the random sequences $(W^{n},X^{n},Y^{n})$ so generated
are almost uniformly distributed over the typical set $\mathcal{T}_{\epsilon}^{n}(Q_{WXY})$;
and on the other hand, the R\'enyi common informations can be expressed
as some R\'enyi divergences. Moreover, these R\'enyi divergences evaluated
at the truncated distribution are almost the same regardless of the
parameter $s\in(-1,1]$. Therefore, by using this truncated code,
Wyner's common information is achievable for any $s\in(-1,1]$.

However, the proof of the converse part for the normalized R\'enyi common
information is not straightforward.\footnote{More precisely, the proof of the converse part for $s\in(-1,0)$ case
is not easy. The converse part for $s=0$ case was proven by Wyner
{[}1{]}, in which the continuity of the normalized entropy under the
normalized KL divergence measure was used, i.e., when $\frac{1}{n}D(P_{X^{n}Y^{n}}\|\pi_{XY}^{n})$
is small then $\frac{1}{n}H_{P}(X^{n}Y^{n})$ is arbitrarily close
to $H_{\pi}(XY)$. But it is not straightforward to apply Wyner's
proof to the case $s\in(-1,0)$, since we do not know whether a strong
enough continuity condition for the normalized entropy holds under
the normalized R\'enyi divergence measure with order \emph{$\alpha=1+s\in(0,1)$}.
Even if a strong enough continuity condition holds, it is not straightforward
to prove. Note that we were not able to directly utilize ideas in
Wyner's converse proof to demonstrate this point. The definition of
the ``relative entropy typical set'' $A(n,\epsilon_{1}):=\big\{\boldsymbol{u}:\frac{1}{n}\log\frac{p_{1}(\boldsymbol{u})}{p_{0}(\boldsymbol{u})}\leq\epsilon_{1}\big\}$
is crucial in Wyner's proof. If we adopt this set with this definition
for the R\'enyi divergence setting (with R\'enyi parameter $<1$), it
is not clear to us whether $\mathbb{P}_{1}(A^{c}(n,\epsilon_{1}))$
vanishes (cf.\ Equation~(A.7) in Wyner's paper).} We attempted to use the method of types to prove it, just as in \cite{Yu}
for the R\'enyi resovability problem, but failed since the code for
the common information problem is arbitrary and does not need to be
i.i.d. In particular, it is not i.i.d. In the following two sections,
we provide an indirect proof using the following strategy: We first
prove an exponential strong converse for Wyner's common information
problem under the TV distance measure in Section \ref{sec:Exponential-Strong-Converse}.
Then by using a relationship between the R\'enyi divergence and the
TV distance \cite{sason2016renyi}, we show this exponential strong
converse implies the converse for normalized R\'enyi divergence in Section
\ref{sec:conversefornormRenyi}.

As an intermediate result, the common information under the TV distance
measure is characterized in the following theorem.
\begin{thm}[Common Information under the TV Distance Measure]
\label{thm:CIunderTV} The following hold:
\begin{enumerate}
\item[(i)] The common information under the TV distance measure satisfies
\begin{equation}
T_{\eps}^{\mathsf{TV}}(\pi_{XY})=\begin{cases}
C_{\mathsf{Wyner}}(X;Y) & \eps\in[0,1)\\
0 & \eps=1
\end{cases}.
\end{equation}
Hence, the strong converse property for the common information problem
under the TV distance holds.
\item[(ii)] Furthermore, there exists a sequence of synthesis codes with rate
$R>C_{\mathsf{Wyner}}(X;Y),$ such that $\left|P_{X^{n}Y^{n}}-\pi_{X^{n}Y^{n}}\right|$
tends to zero exponentially fast as $n$ tends to infinity.
\item[(iii)] On the other hand, for any sequence of synthesis codes with rate
$R<C_{\mathsf{Wyner}}(X;Y),$ we have that $\left|P_{X^{n}Y^{n}}-\pi_{X^{n}Y^{n}}\right|$
tends to one exponentially fast as $n$ tends to infinity.
\end{enumerate}
\end{thm}
Part (ii) is an exponential achievability result while the part (iii)
is an exponential strong converse result. Combining parts (ii) and
(iii) implies part (i). By Pinsker's inequality for R\'enyi divergences~\cite{Erven},
the achievability results (including the exponential achievability
result) in Theorem~\ref{thm:RenyiCI} implies the achievability results
(including the exponential achievability result) in Theorem~\ref{thm:CIunderTV}.
Conversely, the exponential strong converse result in part (iii) of
Theorem \ref{thm:CIunderTV} implies the converse results in Theorem
\ref{thm:RenyiCI} for both unnormalized and normalized R\'enyi divergences.
To prove part (iii), we draw on several key ideas from Oohama's work~\cite{oohama2016exponent}
on the exponential strong converse for the Wyner-Ziv problem. However,
there are several key differences in our proofs, including the way
we establish that the strong converse exponent is positive for all
rates larger than $C_{\mathsf{Wyner}}(X;Y)$ and the treatment of
the cases when various probability mass functions take on the value
zero.

We note that conclusion in part (ii) (the exponential achievability
result) in Theorem \ref{thm:CIunderTV} can be also obtained by using
the soft-covering lemma by Cuff~\cite[Lemma IV.1]{Cuff}.

The proof of the conclusion in part (iii) is provided in the next
section. As mentioned above, the other parts follow directly from
Theorem \ref{thm:RenyiCI}.

\section{\label{sec:Exponential-Strong-Converse} The Proof of Part (iii)
in Theorem \ref{thm:CIunderTV} }

In this section, we provide an exponential strong converse theorem
for the common information problem under the TV distance measure,
which will be used to derive the converse for normalized R\'enyi divergence
in next section.

We define
\begin{align}
\calQ:= & \Big\{ Q_{XYU}\in\calP(\calX\times\calY\times\calU):~\nonumber \\
 & |\calU|\leq|\calX||\calY|,\mathrm{supp}(Q_{XY})\subseteq\mathrm{supp}(\pi_{XY})\Big\}.\label{eq:Q}
\end{align}
Given $\alpha\in[0,1]$ and an arbitrary distribution $Q_{XYU}\in\calQ$,
define the linear combination of the likelihood ratios for $(x,y,u)\in\mathrm{supp}(Q_{XYU})$,
\begin{align}
 & \omega_{Q_{XYU}}^{(\alpha)}(x,y|u):=\bar{\alpha}\Bigg(\log\frac{Q_{XY}(x,y)}{\pi_{XY}(x,y)}\nonumber \\
 & +\log\frac{Q_{XY|U}(x,y|u)}{Q_{X|U}(x|u)Q_{Y|U}(y|u)}\Bigg)+\alpha\log\frac{Q_{XY|U}(x,y|u)}{\pi_{XY}(x,y)}.\label{eq:omega}
\end{align}
This function is finite for all $(x,y,u)\in\mathrm{supp}(Q_{XYU})$.
For $Q_{XYU}\in\calQ$ and $\theta\in[0,\infty)$, define the negative
cumulant generating functions as
\begin{align}
 \Omega^{(\alpha,\theta)}(Q_{XYU})  & \nonumber\\
& \hspace{-.5in}:=-\log\mathbb{E}_{Q_{XYU}}\Big[\exp\big(-\theta\omega_{Q_{XYU}}^{(\alpha)}(X,Y|U)\big)\Big],\label{eq:OmegaQ}
\end{align}
and
\begin{align}
\Omega^{(\alpha,\theta)} & :=\min_{Q_{XYU}\in\calQ}\Omega^{(\alpha,\theta)}(Q_{XYU}),\label{eq:Omega}
\end{align}
where the expectation $\mathbb{E}_{Q_{XYU}}$ is only taken over the set $\mathrm{supp}(Q_{XYU})$
(this means we only sum over the elements $(x,y,u)$ such that $Q_{XYU}(x,y,u)>0$).

Finally, we define the large deviations rate functions
\begin{align}
F^{(\alpha,\theta)}(R) & :=\frac{\Omega^{(\alpha,\theta)}-\theta\alpha R}{1+(5-3\alpha)\theta},\label{def:lossyFalphamubetalambda}\\*
F(R) & :=\sup_{(\alpha,\theta)\in[0,1]\times[0,\infty)}F^{(\alpha,\theta)}(R).\label{def:lossyF}
\end{align}

In view of the definitions above, we have the following theorem. The
proof of this theorem is provided in Appendix~\ref{sec:Proof-of-Theoremexpstrong}.
\begin{thm}
\label{thm:expstrongconverseforTV0} For any synthesis code such that
\begin{align}
\frac{1}{n}\log|{\cal M}_{n}| & \leq R,
\end{align}
we have
\begin{equation}
\left|P_{X^{n}Y^{n}}-\pi_{X^{n}Y^{n}}\right|\geq1-4\exp\big(-nF(R)\big).\label{eq:expbound}
\end{equation}
\end{thm}
If we show $F(R)>0$, then Theorem \ref{thm:expstrongconverseforTV0}
implies the exponential strong converse for TV distance measure. To
that end, we need the following lemma.
\begin{lem}
\label{propFOmega} The following conclusions hold.
\begin{itemize}
\item[(i)] If $R<C_{\mathsf{Wyner}}(X;Y)$, then
\begin{align}
F(R) & >0.
\end{align}
\item[(ii)] If $R\geq C_{\mathsf{Wyner}}(X;Y)$, then
\begin{align}
F(R)=0.
\end{align}
\end{itemize}
\end{lem}
The proof of Lemma \ref{propFOmega} is provided in Appendix \ref{proofpropFOmega}.
We remark that Lemma \ref{propFOmega}, especially part (i), plays
an central role in claiming the exponential strong converse theorem
for the common information problem with the TV distance measure. Its
proof is completely different from that for the corresponding statement
in~\cite{oohama2016exponent} and requires some intricate continuity
arguments (e.g.,~\cite[Lemma~14]{tan2011large}). As we have seen
in Theorem \ref{thm:expstrongconverseforTV0}, $F(R)$ in \eqref{def:lossyF}
is a lower bound on the exponent of $1-\left|P_{X^{n}Y^{n}}-\pi_{X^{n}Y^{n}}\right|$.
This can be regarded as the {\em strong converse exponent}.

Combining Lemma \ref{propFOmega} and Theorem \ref{thm:expstrongconverseforTV0},
we conclude that the exponent in the right hand side of \eqref{eq:expbound}
is strictly positive if the rate is smaller than $C_{\mathsf{Wyner}}(X;Y)$.
Hence, we obtain the exponential strong converse result given in the
conclusion (iii) of Theorem \ref{thm:CIunderTV}.

\section{\label{sec:conversefornormRenyi}Converse Proof of Theorem \ref{thm:RenyiCI}
for the Normalized R\'enyi Common Information }

In this section, we provide a proof of the converse part of Theorem
\ref{thm:RenyiCI} for the normalized R\'enyi common information. To
this end, we need the following relationships between the R\'enyi divergence
and the TV distance.
\begin{lem}[Relationship between the R\'enyi Divergence and the TV Distance (Sason~\cite{sason2016renyi})]
\label{lem:Renyiinequality} For any $s\in(-1,+\infty)$,
\begin{align}
 & \inf_{P_{X},Q_{X}:\left|P_{X}-Q_{X}\right|\geq\epsilon}D_{1+s}(P_{X}\|Q_{X})\nonumber \\
 & =\inf_{P_{X},Q_{X}:\left|P_{X}-Q_{X}\right|=\epsilon}D_{1+s}(P_{X}\|Q_{X})\\
 & =\inf_{q\in[0,1-\epsilon]}d_{1+s}(q+\epsilon\|q),\label{eq:-25}
\end{align}
and for any $s\in(0,1)$,
\begin{align}
 & \inf_{q\in[0,1-\epsilon]}d_{1-s}(q+\epsilon\|q)\nonumber \\
 & \geq\left[\min\left\{ 1,\frac{1-s}{s}\right\} \log\frac{1}{1-\epsilon}-\frac{1}{s}\log2\right]^{+},\label{eq:-27}
\end{align}
where
\begin{equation}
d_{1+s}(p\|q):=\begin{cases}
\frac{1}{s}\log(p^{1+s}q^{-s}+\bar{p}^{1+s}\bar{q}^{-s}), & s\geq-1,s\neq0\\
p\log\frac{p}{q}+\bar{p}\log\frac{\bar{p}}{\bar{q}}, & s=0
\end{cases}
\end{equation}
denotes the binary R\'enyi divergence of order $1+s$.\footnote{For $s=0$ case, the binary R\'enyi divergence is known as the binary
relative entropy, and it is usually denoted as $d(p\|q)$.} We also have
\begin{align}
 & \inf_{P_{X},Q_{X}:\left|P_{X}-Q_{X}\right|\geq\epsilon}D_{0}(P_{X}\|Q_{X})\nonumber \\
 & =\inf_{P_{X},Q_{X}:\left|P_{X}-Q_{X}\right|=\epsilon}D_{0}(P_{X}\|Q_{X})\\
 & =0.
\end{align}
\end{lem}
\begin{rem}
Pinsker's inequality provides a lower bound for $\inf_{P_{X},Q_{X}:\left|P_{X}-Q_{X}\right|\geq\epsilon}\allowdisplaybreaks D_{1+s}(P_{X}\|Q_{X})$
or $\inf_{P_{X},Q_{X}:\left|P_{X}-Q_{X}\right|=\epsilon}$ \allowdisplaybreaks
$D_{1+s}(P_{X}\|Q_{X})$, i.e.,
\begin{align}
 & \inf_{P_{X},Q_{X}:\left|P_{X}-Q_{X}\right|=\epsilon}D_{1+s}(P_{X}\|Q_{X})\geq\frac{(1+s)\epsilon^{2}}{2}.
\end{align}
Hence $\frac{(1+s)\epsilon^{2}}{2}$ is also a lower bound of $\inf_{q\in[0,1-\epsilon]}d_{1+s}(q+\epsilon\|q)$.
\end{rem}
\begin{rem}
Using \eqref{eq:-25} and the lower bound in \eqref{eq:-27}, it is
easy to obtain the following improved lower bounds. For any $s\in(0,1)$,
\begin{align}
 & \inf_{P_{X},Q_{X}:\left|P_{X}-Q_{X}\right|\geq\epsilon}D_{1-s}(P_{X}\|Q_{X})\nonumber \\
 & =\inf_{P_{X},Q_{X}:\left|P_{X}-Q_{X}\right|\geq\epsilon}\sup_{t\in[s,1)}D_{1-t}(P_{X}\|Q_{X})\allowdisplaybreaks\\
 & \geq\sup_{t\in[s,1)}\inf_{P_{X},Q_{X}:\left|P_{X}-Q_{X}\right|\geq\epsilon}D_{1-t}(P_{X}\|Q_{X})\allowdisplaybreaks\\
 & \geq\sup_{t\in[s,1)}\inf_{q\in[0,1-\epsilon]}d_{1-t}(q+\epsilon\|q)\allowdisplaybreaks\\
 & \geq\sup_{t\in[s,1)}\left[\min\left\{ 1,\frac{1-t}{t}\right\} \log\frac{1}{1-\epsilon}-\frac{1}{t}\log2\right]^{+}\allowdisplaybreaks\\
 & =\begin{cases}
\left[\log\frac{1}{4\left(1-\epsilon\right)}\right]^{+} & s\in(0,\frac{1}{2}],\\
\left[\frac{1-s}{s}\log\frac{1}{1-\epsilon}-\frac{1}{s}\log2\right]^{+} & s\in(\frac{1}{2},1),\epsilon>\frac{1}{2}\\
0 & s\in(\frac{1}{2},1),\epsilon\leq\frac{1}{2}
\end{cases}\label{eq:-28}
\end{align}
and for any $s\in[0,+\infty)$,
\begin{align}
 & \inf_{P_{X},Q_{X}:\left|P_{X}-Q_{X}\right|\geq\epsilon}D_{1+s}(P_{X}\|Q_{X})\nonumber \\
 & \geq\inf_{P_{X},Q_{X}:\left|P_{X}-Q_{X}\right|\geq\epsilon}\sup_{t\in(0,1)}D_{1-t}(P_{X}\|Q_{X})\allowdisplaybreaks\\
 & \geq\sup_{t\in(0,1)}\inf_{P_{X},Q_{X}:\left|P_{X}-Q_{X}\right|\geq\epsilon}D_{1-t}(P_{X}\|Q_{X})\allowdisplaybreaks\\
 & \geq\sup_{t\in(0,1)}\inf_{q\in[0,1-\epsilon]}d_{1-t}(q+\epsilon\|q)\allowdisplaybreaks\\
 & \geq\sup_{t\in(0,1)}\left[\min\left\{ 1,\frac{1-t}{t}\right\} \log\frac{1}{1-\epsilon}-\frac{1}{t}\log2\right]^{+}\allowdisplaybreaks\\
 & =\left[\log\frac{1}{4\left(1-\epsilon\right)}\right]^{+}.\label{eq:-29}
\end{align}
\end{rem}
\begin{rem}
The improved lower bounds \eqref{eq:-28} and \eqref{eq:-29} (or
combining \eqref{eq:-25} and the lower bound in \eqref{eq:-27})
implies if
\begin{align}
\left|P_{X}-Q_{X}\right| & \rightarrow1,
\end{align}
then for any $s\in(-1,+\infty)$,
\begin{equation}
D_{1+s}(P_{X}\|Q_{X})\rightarrow\infty.
\end{equation}
\end{rem}
Combining Lemma \ref{lem:Renyiinequality} with Theorem \ref{thm:expstrongconverseforTV0},
we have the converse part for the normalized R\'enyi divergence, which
implies the strong converse for the unnormalized R\'enyi divergence.
\begin{thm}
\label{thm:conversenormRenyi} For any synthesis codes such that
\begin{align}
\limsup_{n\to\infty}\frac{1}{n}\log|{\cal M}_{n}| & <C_{\mathsf{Wyner}}(X;Y),
\end{align}
we have for any $s>-1$,
\begin{equation}
\liminf_{n\to\infty}\frac{1}{n}D_{1+s}(P_{X^{n}Y^{n}}\|\pi_{X^{n}Y^{n}})>0.
\end{equation}
\end{thm}
\begin{rem}
This theorem establishes the converse part of Theorem \ref{thm:RenyiCI}
for the normalized R\'enyi common information.
\end{rem}
\begin{rem}
Since $\liminf_{n\to\infty}\frac{1}{n}D_{1+s}(P_{X^{n}Y^{n}}\|\pi_{X^{n}Y^{n}})>0$
implies $D_{1+s}(P_{X^{n}Y^{n}}\|\pi_{X^{n}Y^{n}})\rightarrow\infty$,
the theorem above implies the strong converse for the Wyner's common
information problem under the unnormalized R\'enyi divergence.
\end{rem}
\begin{IEEEproof}
Theorem \ref{thm:expstrongconverseforTV0} states if $\frac{1}{n}\log|{\cal M}_{n}|<C_{\mathsf{Wyner}}(X;Y)$,
then $\left|P_{X^{n}Y^{n}}-\pi_{X^{n}Y^{n}}\right|\to1$ exponentially
fast. In other words,
\begin{align}
\left|P_{X^{n}Y^{n}}-\pi_{X^{n}Y^{n}}\right| & \geq1-\e^{-n\delta_{n}},
\end{align}
for some sequence $\delta_{n}>0$ such that $\liminf_{n\to\infty}\delta_{n}>0$.
Therefore, using Lemma \ref{lem:Renyiinequality} we have
\begin{align}
 & \liminf_{n\to\infty}\frac{1}{n}D_{1+s}(P_{X^{n}Y^{n}}\|\pi_{X^{n}Y^{n}})\nonumber \\
 & \geq\liminf_{n\to\infty}\left\{ \min\left\{ 1,\frac{1-s}{s}\right\} \delta_{n}-\frac{1}{ns}\log2\right\} \\
 & =\min\left\{ 1,\frac{1-s}{s}\right\} \liminf_{n\to\infty}\delta_{n}\\
 & >0.
\end{align}
This completes the proof.
\end{IEEEproof}

\section{Conclusion and Future Work}

\label{sec:concl}In this paper, we studied a generalized version
of Wyner's common information problem (or the distributed source simulation
problem), in which the unnormalized and normalized R\'enyi divergences
were used to measure the level of approximation. We showed the minimum
rate needed to ensure that the unnormalized or normalized R\'enyi divergence
vanishes asymptotically remains the same as the one under Wyner's
setting where the relative entropy was used.

In the future, we plan to investigate the second-order coding rate
for Wyner's common information under the unnormalized R\'enyi divergence
or the TV distance. For the unnormalized R\'enyi divergence, the one-shot
achievability bound given in Lemma~\ref{lem:oneshotach} can be used
to obtain an achievability bound for the second-order coding rate.
In fact, it can easily be shown that the optimal second-order coding
rate scales as $O(\frac{1}{\sqrt{n}})$. For the TV distance, the
one-shot achievability bound given by Cuff~\cite{Cuff} can be used
to derive an achievability bound. However, the converse parts for
both cases are not straightforward. One may leverage the perturbation
approach~\cite{GuEffros} used to prove the second-order coding rate
for the Gray-Wyner problem in~\cite{watanabe2017second,zhou17}.
This is left as future work.

Furthermore, we are also interested in various closely-related problems.
Among them, the most interesting one is the {\em distributed channel
synthesis problem under the R\'enyi divergence measure}: The coordination
problem or distributed channel synthesis problem was studied
by Cuff, Permuter, and Cover \cite{Cuff10,Cuff}. In this problem,
an observer (encoder) of a source sequence describes the sequence
to a distant random number generator (decoder) that produces another
sequence. What is the minimum description rate needed to produce achieve
a joint distribution that is statistically indistinguishable, under
the TV distance, from the distribution induced by a given channel?
For this problem, Cuff \cite{Cuff} provided a complete characterization
of the minimum rate. We can enhance the level of coordination by replacing
the TV measure with the R\'enyi divergence. For this enhanced version
of the problem, we are interested in characterizing the corresponding
admissible rate region.

\appendices{}

\section{\label{sec:Proof-of-Theoremunnormalized}Achievability Proof of Theorem \ref{thm:RenyiCI}
for the Unnormalized R\'enyi Common Information}

\subsection{Achievability}

\label{sub:Ach}

Next we focus on the achievability part. We first consider the case
$s\in(0,1]$. First we introduce the following one-shot achievability
bound (i.e., achievability bound for blocklength $n$ equal to 1).
\begin{lem}[One-Shot Achievability Bound]
\cite{Yu}\label{lem:oneshotach} Consider a random mapping $P_{X|W}$
and a random codebook $U=\{W(i)\}_{i\in\calM}$ with $W(i)\sim P_{W},i\in\calM$,
where $\calM=\{1,\ldots,\e^{R}\}$. We define
\begin{equation}
P_{X|U}(x|\left\{ w(i)\right\} _{i\in\calM}):=\frac{1}{|{\cal M}|}\sum_{m\in{\cal M}}P_{X|W}(x|w(m))\label{eq:-113-1}
\end{equation}
Then we have for $s\in(0,1]$,
\begin{align}
 & \e^{sD_{1+s}(P_{X|U}\|\pi_{X}|P_{U})}\nonumber \\
 & \leq\e^{sD_{1+s}\left(P_{X|W}\|\pi_{X}|P_{W}\right)-sR}+\e^{sD_{1+s}(P_{X}\|\pi_{X})}\label{eq:-123}\\
 & \leq2\e^{s\Gamma_{1+s}(P_{W},P_{X|W},\pi_{X},R)},\label{eq:-4}
\end{align}
where
\begin{align}
 & \Gamma_{1+s}(P_{W},P_{X|W},\pi_{X},R)\nonumber \\
 & :=\max\left\{ D_{1+s}(P_{X|W}\|\pi_{X}|P_{W})-R,D_{1+s}(P_{X}\|\pi_{X})\right\} .\label{eq:-97}
\end{align}
\end{lem}
\begin{rem}
This lemma provides a one-shot achievability bound for general source
synthesis problems, not only for the distributed source synthesis
or common information problem as studied in this paper.
\end{rem}
By setting $\pi_{X}$, $P_{X|W}$, $P_{W}$, and $R$ to $\pi_{X^{n}Y^{n}}$,
$P_{X^{n}Y^{n}|W^{n}}=P_{X^{n}|W^{n}}P_{Y^{n}|W^{n}}$,\footnote{The pair $(X^{n},Y^{n})$ plays the role of $X$ in Lemma \ref{lem:oneshotach}.}
$P_{W^{n}}$, and $nR$ respectively, Lemma \ref{lem:oneshotach}
can be used to derive an achievability result for the common information
problem. Applying Lemma \ref{lem:oneshotach} and taking limits appropriately,
we obtain if there exists a sequence of distributions $\left\{ P_{W^{n}}P_{X^{n}|W^{n}}P_{Y^{n}|W^{n}}\right\} $
such that $\lim_{n\to\infty}D_{1+s}(P_{X^{n}Y^{n}}\|\pi_{X^{n}Y^{n}})\rightarrow0$
and $R>\limsup_{n\to\infty}\frac{1}{n}D_{1+s}(P_{X^{n}Y^{n}|W^{n}}\|\pi_{X^{n}Y^{n}}|P_{W^{n}})$,
then there exists a sequence of codes such that
\begin{align}
 & \limsup_{n\to\infty}D_{1+s}(P_{X^{n}Y^{n}|U_{n}}\|\pi_{X^{n}Y^{n}}|P_{U_{n}})\nonumber \\
 & \leq\limsup_{n\to\infty}\frac{1}{s}\log\Bigl\{\e^{sD_{1+s}\left(P_{X^{n}Y^{n}|W^{n}}\|\pi_{X^{n}Y^{n}}|P_{W^{n}}\right)-nsR}\nonumber \\
 & \qquad+\e^{sD_{1+s}(P_{X^{n}Y^{n}}\|\pi_{X^{n}Y^{n}})}\Bigr\}\\
 & \leq\frac{1}{s}\log\Bigl\{\limsup_{n\to\infty}\e^{s\left(D_{1+s}\left(P_{X^{n}Y^{n}|W^{n}}\|\pi_{X^{n}Y^{n}}|P_{W^{n}}\right)-nR\right)}+1\Bigr\}\\
 & \leq\frac{1}{s}\log\Bigl\{\limsup_{n\to\infty}\e^{s\left(n(R-\epsilon)-nR\right)}+1\Bigr\}\label{eq:-5}\\
 & =0,
\end{align}
where \eqref{eq:-5} follows since
\[
R>\limsup_{n\to\infty}\frac{1}{n}D_{1+s}(P_{X^{n}Y^{n}|W^{n}}\|\pi_{X^{n}Y^{n}}|P_{W^{n}})
\]
 implies there exists a constant $\epsilon>0$ such that
\[
R-\epsilon>\frac{1}{n}D_{1+s}(P_{X^{n}Y^{n}|W^{n}}\|\pi_{X^{n}Y^{n}}|P_{W^{n}})
\]
 holds for all sufficiently large $n$. Therefore, the minimum achievable
rate satisfies
\begin{align}
 & \inf\left\{ R:D_{1+s}(P_{X^{n}Y^{n}|U_{n}}\|\pi_{X^{n}Y^{n}}|P_{U_{n}})\rightarrow0\right\} \nonumber \\
\leq & \inf_{\substack{\{P_{W^{n}},P_{X^{n}|W^{n}},P_{Y^{n}|W^{n}}\}_{n=1}^{\infty}:\\
D_{1+s}(P_{X^{n}Y^{n}}\|\pi_{X^{n}Y^{n}})\rightarrow0
}
}\nonumber \\
 & \qquad\limsup_{n\to\infty}\frac{1}{n}D_{1+s}(P_{X^{n}Y^{n}|W^{n}}\|\pi_{X^{n}Y^{n}}|P_{W^{n}}).\label{eq:-3}
\end{align}

Let $Q_{WXY}$ be a distribution such that $Q_{XY}=\pi_{XY}$ and
$X-W-Y$. For the optimization in \eqref{eq:-3}, to obtain an upper
bound, we set the distributions
\begin{align*}
P_{W^{n}}\left(w^{n}\right) & \propto Q_{W}^{n}\left(w^{n}\right)1\left\{ w^{n}\in\mathcal{T}_{\epsilon'}^{n}\left(Q_{W}\right)\right\} ,\\
P_{X^{n}|W^{n}}\left(x^{n}|w^{n}\right) & \propto Q_{X|W}^{n}\left(x^{n}|w^{n}\right)1\left\{ x^{n}\in\mathcal{T}_{\epsilon}^{n}\left(Q_{WX}|w^{n}\right)\right\} ,\\
P_{Y^{n}|W^{n}}\left(x^{n}|w^{n}\right) & \propto Q_{Y|W}^{n}\left(x^{n}|w^{n}\right)1\left\{ y^{n}\in\mathcal{T}_{\epsilon}^{n}\left(Q_{WY}|w^{n}\right)\right\} ,
\end{align*}
where $0<\epsilon'<\epsilon\leq1$. Then we have
\begin{align}
 & P_{X^{n}Y^{n}}\left(x^{n},y^{n}\right)\nonumber \\
 & =\sum_{w^{n}}P_{W^{n}}\left(w^{n}\right)P_{X^{n}|W^{n}}\left(x^{n}|w^{n}\right)P_{Y^{n}|W^{n}}\left(x^{n}|w^{n}\right)\\
 & =\sum_{w^{n}}\frac{Q_{W}^{n}\left(w^{n}\right)1\left\{ w^{n}\in\mathcal{T}_{\epsilon'}^{n}\left(Q_{W}\right)\right\} }{Q_{W}^{n}\left(\mathcal{T}_{\epsilon'}^{n}\right)}\nonumber \\
 & \qquad\times\frac{Q_{X|W}^{n}\left(x^{n}|w^{n}\right)1\left\{ x^{n}\in\mathcal{T}_{\epsilon}^{n}\left(Q_{WX}|w^{n}\right)\right\} }{Q_{X|W}^{n}\left(\mathcal{T}_{\epsilon}^{n}\left(Q_{WX}|w^{n}\right)|w^{n}\right)}\nonumber \\
 & \qquad\times\frac{Q_{Y|W}^{n}\left(x^{n}|w^{n}\right)1\left\{ y^{n}\in\mathcal{T}_{\epsilon}^{n}\left(Q_{WY}|w^{n}\right)\right\} }{Q_{Y|W}^{n}\left(\mathcal{T}_{\epsilon}^{n}\left(Q_{WY}|w^{n}\right)|w^{n}\right)}\\
 & \leq\frac{\sum_{w^{n}}Q_{WXY}^{n}\left(w^{n},x^{n},y^{n}\right)}{Q_{W}^{n}\left(\mathcal{T}_{\epsilon'}^{n}\right)}\nonumber \\
 & \qquad\times\frac{1}{\min_{w^{n}\in\mathcal{T}_{\epsilon'}^{n}}Q_{X|W}^{n}\left(\mathcal{T}_{\epsilon}^{n}\left(Q_{WX}|w^{n}\right)|w^{n}\right)}\nonumber \\
 & \qquad\times\frac{1}{\min_{w^{n}\in\mathcal{T}_{\epsilon'}^{n}}Q_{Y|W}^{n}\left(\mathcal{T}_{\epsilon}^{n}\left(Q_{WY}|w^{n}\right)|w^{n}\right)}\label{eq:-23}\\
 & =\frac{\pi_{X^{n}Y^{n}}(x^{n},y^{n})}{1-\delta_{n}},\label{eq:-12}
\end{align}
where in \eqref{eq:-12} $\delta_{n}$ is defined as $1$ minus the
denominator of \eqref{eq:-23}. Here we claim that $\delta_{n}\to0$
as $n\rightarrow\infty$. This follows since $Q_{W}^{n}(\mathcal{T}_{\epsilon'}^{n})\rightarrow1,$
$\min_{w^{n}\in\mathcal{T}_{\epsilon'}^{n}}Q_{X|W}^{n}\bigl(\mathcal{T}_{\epsilon}^{n}(Q_{WX}|w^{n})|w^{n}\bigr)\rightarrow1,$
and $\min_{w^{n}\in\mathcal{T}_{\epsilon'}^{n}}Q_{Y|W}^{n}\bigl(\mathcal{T}_{\epsilon}^{n}(Q_{WY}|w^{n})|w^{n}\bigr)\rightarrow1$,
where the last two limits hold due to the following lemma.
\begin{lem}
\label{lem:contyplem} Assume $0<\epsilon'<\epsilon\leq1$, then as
$n\to\infty$, $Q_{X|W}^{n}\bigl(\mathcal{T}_{\epsilon}^{n}(Q_{WX}|w^{n})|w^{n}\bigr)$
converges uniformly\footnote{This means that for any $\eta>0$, there exists an integer $N=N_{\eta}$
such that $\max_{w^{n}\in\calT_{\epsilon'}^{n}(Q_{W})}1-Q_{X|W}^{n}\bigl(\mathcal{T}_{\epsilon}^{n}(Q_{WX}|w^{n})|w^{n}\bigr)\le\eta$
for all $n>N_{\eta}$. Here the notion of ``uniform convergence''
is a slightly different from the conventional one~\cite[Definition~7.7]{Rudin}.
In the conventional definition, the domain of the functions are fixed
but here, the domain $\calT_{\epsilon'}^{n}(Q_{W})$ depends on~$n$. } to $1$ (in $w^{n}\in\mathcal{T}_{\epsilon'}^{n}\left(Q_{W}\right)$).
\begin{align}
 & 1-Q_{X|W}^{n}\bigl(\mathcal{T}_{\epsilon}^{n}(Q_{WX}|w^{n})|w^{n}\bigr)\leq\nonumber \\
 & \left|\mathcal{X}\right|\left|\mathcal{W}\right|\Big(\e^{-\frac{1}{3}\left(\frac{\epsilon-\epsilon'}{1+\epsilon'}\right)^{2}nQ_{X|W}^{(\mathsf{min})}}+\e^{-\frac{1}{2}\left(\frac{\epsilon-\epsilon'}{1-\epsilon'}\right)^{2}nQ_{X|W}^{(\mathsf{min})}}\Big),
\end{align}
for all $w^{n}\in\mathcal{T}_{\epsilon'}^{n}\left(Q_{W}\right)$,
where $Q_{X|W}^{(\mathsf{min})}:=\min_{(x,w):Q_{X|W}(x|w)>0}Q_{X|W}(x|w)$.
\end{lem}
This lemma is a stronger version of the conditional typicality lemma
in \cite{Gamal}, since here the probability converges uniformly,
instead of converging pointwise. However, the proof is merely a refinement
of the conditional typicality lemma \cite[Appendix~2A]{Gamal} (by
applying the Chernoff bound, instead of the law of large numbers),
and hence omitted here. Besides, a similar lemma can be found in \cite[Lemma 2.12]{Csiszar},
which is established based on a slightly different definition of strong
typicality.

Using this upper bound of $P_{X^{n}Y^{n}}(x^{n},y^{n})$ we have
\begin{align}
 & D_{1+s}(P_{X^{n}Y^{n}}\|\pi_{X^{n}Y^{n}})\nonumber \\
 & =\frac{1}{s}\log\sum_{x^{n},y^{n}}P_{X^{n}Y^{n}}^{1+s}(x^{n},y^{n})\pi_{X^{n}Y^{n}}^{-s}(x^{n},y^{n})\\
 & \leq\frac{1}{s}\log\sum_{x^{n},y^{n}}\left(\frac{\pi_{X^{n}Y^{n}}(x^{n},y^{n})}{1-\delta_{n}}\right)^{1+s}\pi_{X^{n}Y^{n}}^{-s}(x^{n},y^{n})\\
 & =\frac{1}{s}\log\left(\frac{1}{1-\delta_{n}}\right)^{1+s}\\
 & \rightarrow0.\label{eq:-26}
\end{align}

Let $[T_{W}V_{X|W}]$ denote the joint distribution of $X$ and $W$
induced by the type $T_{W}$ and conditional type $V_{X|W}$. Now
define the sets of tuples of types and conditional types:
\begin{align}
 & \calA:=\Bigl\{(T_{W},V_{X|W},V_{Y|W}):\nonumber \\
 & \forall w,\left|T_{W}(w)-Q_{W}(w)\right|\leq\epsilon'Q_{W}(w),\nonumber \\
 & \forall(w,x),\left|[T_{W}V_{X|W}](w,x)-Q_{WX}(w,x)\right|\leq\epsilon Q_{WX}(w,x),\nonumber \\
 & \forall(w,y),\left|[T_{W}V_{Y|W}](w,y)-Q_{WY}(w,y)\right|\leq\epsilon Q_{WY}(w,y)\Bigr\}
\end{align}
and
\begin{align}
 & {\cal B}:=\Bigl\{(T_{W},V_{X|W},V_{Y|W}):\forall(w,x,y),\nonumber \\
 & \frac{(1-\epsilon)^{2}}{1+\epsilon'}\leq\frac{[T_{W}V_{X|W}V_{Y|W}](w,x,y)}{Q_{WXY}(w,x,y)}\leq\frac{(1+\epsilon)^{2}}{1-\epsilon'}\Bigr\}.\label{eqn:setB}
\end{align}
In \eqref{eqn:setB}, if $Q_{WXY}(w,x,y)=0$, this imposes that $\left[T_{W}V_{X|W}V_{Y|W}\right](w,x,y)=0$.
It is easy to verify that ${\cal A}\subseteq{\cal B}$. Let $\delta_{1,n}$
and $\delta_{2,n}$ be two arbitrary sequences tending to zero as
$n\rightarrow\infty$. Using these notations, we can write \eqref{eq:-24}-\eqref{eq:-16}
(shown at the top of the next page),
\begin{figure*}[!t]
\setcounter{mytempeqncnt}{\value{equation}} \setcounter{equation}{75}
\begin{align}
 & \frac{1}{n}D_{1+s}\left(P_{W^{n}X^{n}Y^{n}}\|P_{W^{n}}\pi_{X^{n}Y^{n}}\right)\nonumber \\
 & =\frac{1}{ns}\log\sum_{w^{n},x^{n},y^{n}}P\left(w^{n}\right)\left(P\left(x^{n}|w^{n}\right)P\left(y^{n}|w^{n}\right)\right)^{1+s}\pi^{-s}(x^{n},y^{n})\label{eq:-24}\\
 & =\frac{1}{ns}\log\sum_{w^{n},x^{n},y^{n}}P\left(w^{n},x^{n},y^{n}\right)\nonumber \\
 & \qquad\times\Biggl(\frac{Q_{X|W}^{n}\left(x^{n}|w^{n}\right)1\left\{ x^{n}\in\mathcal{T}_{\epsilon}^{n}\left(Q_{WX}|w^{n}\right)\right\} }{Q_{X|W}^{n}\left(\mathcal{T}_{\epsilon}^{n}\left(Q_{WX}|w^{n}\right)|w^{n}\right)}\frac{Q_{Y|W}^{n}\left(x^{n}|w^{n}\right)1\left\{ y^{n}\in\mathcal{T}_{\epsilon}^{n}\left(Q_{WY}|w^{n}\right)\right\} }{Q_{Y|W}^{n}\left(\mathcal{T}_{\epsilon}^{n}\left(Q_{WY}|w^{n}\right)|w^{n}\right)}\Biggr)^{s}\pi_{X^{n}Y^{n}}^{-s}(x^{n},y^{n})\\
 & =\frac{1}{ns}\log\sum_{T_{W},V_{X|W},V_{Y|W}}\sum_{\substack{\substack{w^{n}\in\mathcal{T}_{T_{W}},x^{n}\in\mathcal{T}_{V_{X|W}}\left(w^{n}\right),\\
y^{n}\in\mathcal{T}_{V_{Y|W}}\left(w^{n}\right)
}
}
}P\left(w^{n},x^{n},y^{n}\right)\nonumber \\*
 & \qquad\times\left(\frac{Q_{X|W}^{n}\left(x^{n}|w^{n}\right)1\left\{ x^{n}\in\mathcal{T}_{\epsilon}^{n}\left(Q_{WX}|w^{n}\right)\right\} }{Q_{X|W}^{n}\left(\mathcal{T}_{\epsilon}^{n}\left(Q_{WX}|w^{n}\right)|w^{n}\right)}\frac{Q_{Y|W}^{n}\left(x^{n}|w^{n}\right)1\left\{ y^{n}\in\mathcal{T}_{\epsilon}^{n}\left(Q_{WY}|w^{n}\right)\right\} }{Q_{Y|W}^{n}\left(\mathcal{T}_{\epsilon}^{n}\left(Q_{WY}|w^{n}\right)|w^{n}\right)}\right)^{s}\pi_{X^{n}Y^{n}}^{-s}(x^{n},y^{n})\\
 & \leq\frac{1}{ns}\log\sum_{(T_{X},V_{X|W},V_{Y|W})\in{\cal A}}\sum_{\substack{\substack{w^{n}\in\mathcal{T}_{T_{W}},x^{n}\in\mathcal{T}_{V_{X|W}}\left(w^{n}\right),\\
y^{n}\in\mathcal{T}_{V_{Y|W}}\left(w^{n}\right)
}
}
}P\left(w^{n},x^{n},y^{n}\right)\nonumber \\
 & \qquad\times\left(\frac{\e^{n\sum_{w,x}T\left(w,x\right)\log Q\left(x|w\right)}}{1-\delta_{1,n}}\frac{\e^{n\sum_{w,y}T\left(w,y\right)\log Q\left(y|w\right)}}{1-\delta_{2,n}}\right)^{s}e^{-ns\sum_{x,y}T\left(x,y\right)\log\pi\left(x,y\right)}\label{eq:}\\
 & \leq-\frac{1}{n}\log(1-\delta_{1,n})(1-\delta_{2,n})+\frac{1}{ns}\log\sum_{(T_{X},V_{X|W},V_{Y|W})\in{\cal B}}\sum_{\substack{\substack{w^{n}\in\mathcal{T}_{T_{W}},x^{n}\in\mathcal{T}_{V_{X|W}}\left(w^{n}\right),\\
y^{n}\in\mathcal{T}_{V_{Y|W}}\left(w^{n}\right)
}
}
}P\left(w^{n},x^{n},y^{n}\right)\nonumber \\
 & \qquad\times\max_{(T_{X},V_{X|W},V_{Y|W})\in{\cal B}}\e^{sn\sum_{w,x}T\left(w,x\right)\log Q\left(x|w\right)+sn\sum_{w,y}T\left(w,y\right)\log Q\left(y|w\right)-ns\sum_{x,y}T\left(x,y\right)\log\pi\left(x,y\right)}\label{eq:-10}\\
 & \leq\max_{(T_{X},V_{X|W},V_{Y|W})\in{\cal B}}\left(\sum_{w,x}T\left(w,x\right)\log Q\left(x|w\right)+\sum_{w,y}T\left(w,y\right)\log Q\left(y|w\right)-\sum_{x,y}T\left(x,y\right)\log\pi\left(x,y\right)\right)\nonumber \\
 & \qquad-\frac{1}{n}\log(1-\delta_{1,n})(1-\delta_{2,n})\label{eq:-9}\\
 & \leq\frac{(1-\epsilon)^{2}}{1+\epsilon'}\left(\sum_{w,x}Q\left(w,x\right)\log Q\left(x|w\right)+\sum_{w,y}Q\left(w,y\right)\log Q\left(y|w\right)\right)-\frac{(1+\epsilon)^{2}}{1-\epsilon'}\sum_{x,y}Q\left(x,y\right)\log\pi\left(x,y\right)\nonumber \\
 & \qquad-\frac{1}{n}\log(1-\delta_{1,n})(1-\delta_{2,n})\label{eq:-13}\\
 & =-\frac{(1-\epsilon)^{2}}{1+\epsilon'}\left(H_{Q}(X|W)+H_{Q}(Y|W)\right)+\frac{(1+\epsilon)^{2}}{1-\epsilon'}H_{Q}(XY)-\frac{1}{n}\log(1-\delta_{1,n})(1-\delta_{2,n})\label{eq:-14}\\
 & =\frac{(1-\epsilon)^{2}}{1+\epsilon'}I_{Q}(XY;W)+\frac{4\epsilon}{1-\epsilon'}H_{Q}(XY)-\frac{1}{n}\log(1-\delta_{1,n})(1-\delta_{2,n}),\label{eq:-16}
\end{align}
\setcounter{mytempeqncnt}{\value{equation}} \setcounter{equation}{\value{mytempeqncnt}}
\hrulefill{}
\end{figure*}
where \eqref{eq:} follows from Lemma \ref{lem:contyplem}, \eqref{eq:-10}
follows since ${\cal A}\subseteq{\cal B}$, \eqref{eq:-9} follows
since
\begin{equation}
\sum_{(T_{X},V_{X|W},V_{Y|W})\in{\cal B}}\sum_{\substack{\substack{w^{n}\in\mathcal{T}_{T_{W}},x^{n}\in\mathcal{T}_{V_{X|W}}\left(w^{n}\right),\\
y^{n}\in\mathcal{T}_{V_{Y|W}}\left(w^{n}\right)
}
}
}P\left(w^{n},x^{n},y^{n}\right)\leq1,\label{eqn:89}
\end{equation}
and \eqref{eq:-14} follows since $Q_{XY}=\pi_{XY}$.

Letting $n\rightarrow\infty$ in \eqref{eq:-16}, we have
\begin{align}
 & \limsup_{n\to\infty}\frac{1}{n}D_{1+s}\left(P_{W^{n}X^{n}Y^{n}}\|P_{W^{n}}\pi_{X^{n}Y^{n}}\right)\nonumber \\
 & \leq\frac{(1-\epsilon)^{2}}{1+\epsilon'}I_{Q}(XY;W)+\frac{4\epsilon}{1-\epsilon'}H_{Q}(XY).\label{eq:-15}
\end{align}
Combining \eqref{eq:-15} with \eqref{eq:-3} and \eqref{eq:-26},
we obtain
\begin{align}
 & \inf\left\{ R:D_{1+s}(P_{X^{n}Y^{n}|U_{n}}\|\pi_{X^{n}Y^{n}}|P_{U_{n}})\rightarrow0\right\} \nonumber \\
 & \leq\frac{(1-\epsilon)^{2}}{1+\epsilon'}I_{Q}(XY;W)+\frac{4\epsilon}{1-\epsilon'}H_{Q}(XY).
\end{align}
Since $\epsilon>\epsilon'>0$ are arbitrary, and $H_{Q}(XY)=H_{\pi}(XY)\leq\log\left\{ |\mathcal{X}||\mathcal{Y}|\right\} $
is bounded, we have
\begin{align}
 & \inf\left\{ R:D_{1+s}(P_{X^{n}Y^{n}|U_{n}}\|\pi_{X^{n}Y^{n}}|P_{U_{n}})\rightarrow0\right\} \nonumber \\
 & \leq I_{Q}(XY;W).
\end{align}
Since the distribution $Q_{WXY}$ is arbitrary, we can minimize $I_{Q}(XY;W)$
over all distributions satisfying $Q_{XY}=\pi_{XY}$ and $X-W-Y$.
Hence
\begin{align}
 & \inf\left\{ R:D_{1+s}(P_{X^{n}Y^{n}|U_{n}}\|\pi_{X^{n}Y^{n}}|P_{U_{n}})\rightarrow0\right\} \nonumber \\
 & \leq\min_{Q_{XYW}:\,Q_{XY}=\pi_{XY},\,X-W-Y}I_{Q}(XY;W)\\
 & =C_{\mathsf{Wyner}}(X;Y).
\end{align}
Observe that
\begin{align}
 & D_{1+s}(P_{X^{n}Y^{n}|U_{n}}\|\pi_{X^{n}Y^{n}}|P_{U_{n}})\nonumber \\
 & =\frac{1}{s}\log\mathbb{E}_{U_{n}}\bigg[\sum_{x^{n},y^{n}}P_{X^{n}Y^{n}|U_{n}}(x^{n},y^{n}|U_{n})\nonumber \\
 & \qquad\times\left(\frac{P_{X^{n}Y^{n}|U_{n}}(x^{n},y^{n}|U_{n})}{\pi_{X^{n}Y^{n}}(x^{n},y^{n})}\right)^{s}\bigg],
\end{align}
where $\mathbb{E}_{U_{n}}$ is the expectation taken with respect
to the distribution $P_{U_{n}}$. Hence $D_{1+s}(P_{X^{n}Y^{n}|U_{n}}\|\pi_{X^{n}Y^{n}}|P_{U_{n}})\rightarrow0$
implies that there must exist at least one sequence of codebooks indexed
by $\{u_{n}\}_{n=1}^{\infty}$ such that $D_{1+s}(P_{X^{n}Y^{n}|U_{n}=u_{n}}\|\pi_{X^{n}Y^{n}})\rightarrow0$.
Therefore, the R\'enyi common information for $s\in(0,1]$ is not larger
than $C_{\mathsf{Wyner}}(X;Y)$. This completes the proof for the
case $s\in(0,1]$.

Now we prove the case $s\in(-1,0)$. Since $D_{1+s}(P_{X^{n}Y^{n}}\|\pi_{X^{n}Y^{n}})$
is non-decreasing in $s$, the result for $s\in(0,1]$ implies the
achievability result for $s\in(-1,0)$.

\subsection{Exponential Achievability}

Since $D_{1+s}(P_{X^{n}Y^{n}U_{n}}\|\pi_{X^{n}Y^{n}}\times P_{U_{n}})$
is non-decreasing in $s$, to prove the exponential result for $s\in(-1,1]$,
we only need to show the result holds for $s\in(0,1]$. To this end,
we use the random code given in Appendix \ref{sub:Ach}. For this
code, by Lemma \ref{lem:oneshotach}, we obtain
\begin{align}
 & \e^{sD_{1+s}(P_{X^{n}Y^{n}U_{n}}\|\pi_{X^{n}Y^{n}}\times P_{U_{n}})}\nonumber \\
 & \leq\e^{sD_{1+s}(P_{W^{n}X^{n}Y^{n}}\|P_{W^{n}}\pi_{X^{n}Y^{n}})-nsR}\nonumber \\
 & \qquad+\e^{sD_{1+s}(P_{X^{n}Y^{n}}\|\pi_{X^{n}Y^{n}})}\\
 & =\e^{sD_{1+s}(P_{X^{n}Y^{n}}\|\pi_{X^{n}Y^{n}})}\bigl(\e^{sD_{1+s}(P_{W^{n}X^{n}Y^{n}}\|P_{W^{n}}\pi_{X^{n}Y^{n}})-nsR}\nonumber \\
 & \qquad\times\e^{-sD_{1+s}(P_{X^{n}Y^{n}}\|\pi_{X^{n}Y^{n}})}+1\bigr).\label{eq:-50-1}
\end{align}
Taking $\log$'s and normalizing by $s$,
\begin{align}
 & D_{1+s}(P_{X^{n}Y^{n}U_{n}}\|\pi_{X^{n}Y^{n}}\times P_{U_{n}})\nonumber \\
 & =D_{1+s}(P_{X^{n}Y^{n}}\|\pi_{X^{n}Y^{n}})\nonumber \\
 & \qquad+\frac{1}{s}\log\Big(\e^{sD_{1+s}(P_{W^{n}X^{n}Y^{n}}\|P_{W^{n}}\pi_{X^{n}Y^{n}})-nsR}\nonumber \\
 & \qquad\times\e^{-sD_{1+s}(P_{X^{n}Y^{n}}\|\pi_{X^{n}Y^{n}})}+1\Big)\\
 & \leq D_{1+s}(P_{X^{n}Y^{n}}\|\pi_{X^{n}Y^{n}})+\frac{1}{s}\e^{sD_{1+s}(P_{W^{n}X^{n}Y^{n}}\|P_{W^{n}}\pi_{X^{n}Y^{n}})}\nonumber \\
 & \qquad\times\e^{-nsR-sD_{1+s}(P_{X^{n}Y^{n}}\|\pi_{X^{n}Y^{n}})}.\label{eq:-54-1}
\end{align}

We first consider the first term of \eqref{eq:-54-1}. Note that in
\eqref{eq:-12}, $\delta_{n}$ tends to zero exponentially fast as
$n\rightarrow\infty$, since $Q_{W}^{n}(\mathcal{T}_{\epsilon'}^{n}),$
$\min_{w^{n}\in\mathcal{T}_{\epsilon'}^{n}}Q_{X|W}^{n}\bigl(\mathcal{T}_{\epsilon}^{n}(Q_{WX}|w^{n})|w^{n}\bigr),$
and $\min_{w^{n}\in\mathcal{T}_{\epsilon'}^{n}}Q_{Y|W}^{n}\bigl(\mathcal{T}_{\epsilon}^{n}(Q_{WY}|w^{n})|w^{n}\bigr)$
all tend to one exponentially fast as $n\rightarrow\infty$. Combining
this with \eqref{eq:-26}, we obtain that $D_{1+s}(P_{X^{n}Y^{n}}\|\pi_{X^{n}Y^{n}})\to0$
exponentially fast.

Furthermore, by \eqref{eq:-16} we can write the exponent of the second
term of \eqref{eq:-54-1} as
\begin{align}
 & \liminf_{n\to\infty}sR-\frac{1}{n}sD_{1+s}(P_{W^{n}X^{n}Y^{n}}\|P_{W^{n}}\pi_{X^{n}Y^{n}})\nonumber \\
 & \qquad+\frac{1}{n}sD_{1+s}(P_{X^{n}Y^{n}}\|\pi_{X^{n}Y^{n}})\nonumber \\
 & =sR-s\left(\frac{(1-\epsilon)^{2}}{1+\epsilon'}I_{Q}(XY;W)+\frac{4\epsilon}{1-\epsilon'}H_{Q}(XY)\right).
\end{align}
Since $H_{Q}(XY)=H_{\pi}(XY)\leq\log\left\{ |\mathcal{X}||\mathcal{Y}|\right\} $
is bounded and $R>I_{Q}(XY;W)$, by choosing sufficiently small $\epsilon>\epsilon'>0$,
we can ensure this exponent is positive.

Combining the two points above, we conclude that the optimal $D_{1+s}(P_{X^{n}Y^{n}U_{n}}\|\pi_{X^{n}Y^{n}}\times P_{U_{n}})$
tends to zero exponentially fast as long as $R>C_{\mathsf{Wyner}}(X;Y)$.
On the other hand, by a similar argument in Appendix \ref{sub:Ach},
$D_{1+s}(P_{X^{n}Y^{n}|U_{n}}\|\pi_{X^{n}Y^{n}}|P_{U_{n}})\rightarrow0$
exponentially fast implies that there must exist at least one sequence
of codebooks indexed by $\{u_{n}\}_{n=1}^{\infty}$ such that
\[
D_{1+s}(P_{X^{n}Y^{n}|U_{n}=u_{n}}\|\pi_{X^{n}Y^{n}})\rightarrow0
\]
exponentially fast. Hence the proof is completed.

\section{\label{sec:Proof-of-Theoremexpstrong}Proof of Theorem \ref{thm:expstrongconverseforTV0}}

\subsection{\label{subsec:Proof-of-Theoremexpstrong}Proof of Theorem \ref{thm:expstrongconverseforTV0}}

In this section, we present the proof of Theorem \ref{thm:expstrongconverseforTV0}.
In the proof, we adapt the information spectrum method proposed by
Oohama~\cite{oohama2016exponent} to first establish a non-asymptotic
lower bound on $\left|P_{X^{n}Y^{n}}-\pi_{X^{n}Y^{n}}\right|$. Invoking
the lower bound (cf.~Lemma~\ref{fblidlossy2}) and applying Cram\'er's
bound in the theory of large deviations~\cite{Dembo}, we can obtain
a further lower bound on $\left|P_{X^{n}Y^{n}}-\pi_{X^{n}Y^{n}}\right|$
leading to \eqref{eq:expbound}.

Let $P_{M_{n}X^{n}Y^{n}}$ be the joint distribution of $(M_{n},X^{n},Y^{n})$,
induced by the synthesis code, i.e.,
\begin{align}
 & P_{M_{n}X^{n}Y^{n}}(m,x^{n},y^{n})\nonumber \\
 & =\frac{1}{|{\cal M}_{n}|}P_{X^{n}|M_{n}}(x^{n}|m)P_{Y^{n}|M_{n}}(y^{n}|m).
\end{align}
In the following, for brevity sometimes we omit the subscript, and
write $P_{M_{n}X^{n}Y^{n}}$ as $P$.

Let $Q_{X^{n}Y^{n}}$ and $Q_{X^{n}Y^{n}|M_{n}}$ be arbitrary distributions.
Given any $\eta>0$, define the following information-spectrum sets
and support sets:
\begin{align}
\calA_{1} & :=\Big\{(x^{n},y^{n}):\nonumber \\
 & \quad\frac{1}{n}\log\frac{\pi_{X^{n}Y^{n}}(x^{n},y^{n})}{Q_{X^{n}Y^{n}}(x^{n},y^{n})}\geq-\eta\Big\}\times{\cal M}_{n},\label{eq:A1}\\
\calA_{2} & :=\Big\{(x^{n},y^{n},m):\nonumber \\
 & \quad\frac{1}{n}\log\frac{P_{X^{n}|M_{n}}(x^{n}|w)P_{Y^{n}|M_{n}}(y^{n}|m)}{Q_{X^{n}Y^{n}|M_{n}}(x^{n},y^{n}|m)}\geq-\eta\Big\},\\
\calA_{3} & :=\Big\{(x^{n},y^{n},m):\nonumber \\
 & \quad\frac{1}{n}\log\frac{Q_{X^{n}Y^{n}|M_{n}}(x^{n},y^{n}|m)}{\pi_{X^{n}Y^{n}}(x^{n},y^{n})}\leq R+\eta\Big\},\label{eq:A3}\\
\widetilde{\calA}_{1} & :=\mathrm{supp}(\pi_{X^{n}Y^{n}})\times{\cal M}_{n},\\
\widetilde{\calA}_{2} & :=\mathrm{supp}(P_{X^{n}Y^{n}M_{n}}),\\
\widetilde{\calA} & :=\widetilde{\calA}_{1}\cap\widetilde{\calA}_{2}.\label{eq:Atilted}
\end{align}

Choose $U_{i}=M_{n}$ and $V_{i}=(X^{i-1},Y^{i-1})$. For $i=1,\ldots,n$,
let $Q_{X_{i}Y_{i}U_{i}V_{i}}$ be any distribution and let $Q_{X^{n}Y^{n}}=\prod_{i=1}^{n}Q_{X_{i}Y_{i}|X^{i-1}Y^{i-1}}=\prod_{i=1}^{n}Q_{X_{i}Y_{i}|V_{i}}$
and $Q_{X^{n}Y^{n}|M_{n}}=\prod_{i=1}^{n}Q_{X_{i}Y_{i}|M_{n}X^{i-1}Y^{i-1}}=\prod_{i=1}^{n}Q_{X_{i}Y_{i}|U_{i}V_{i}}$,
where $Q_{X_{i}Y_{i}|V_{i}}$ and $Q_{X_{i}Y_{i}|U_{i}V_{i}}$ are
conditional distributions induced by $Q_{X_{i}Y_{i}U_{i}V_{i}}$.
Paralleling \eqref{eq:A1} to \eqref{eq:A3}, given any $\eta>0$,
we define the following{} memoryless version of information-spectrum
sets:
\begin{align}
\calB_{1}:= & \Big\{(x^{n},y^{n},v^{n}):\nonumber \\
 & \frac{1}{n}\sum_{i=1}^{n}\log\frac{Q_{X_{i}Y_{i}|V_{i}}(x_{i},y_{i}|v_{i})}{\pi_{XY}(x_{i},y_{i})}\leq\eta\Big\}\times{\cal M}_{n}^{n},\label{eq:B1}\\
\calB_{2}:= & \Big\{(x^{n},y^{n},u^{n},v^{n}):\nonumber \\
 & \frac{1}{n}\sum_{i=1}^{n}\log\frac{Q_{X_{i}Y_{i}|U_{i}V_{i}}(x_{i},y_{i}|u_{i},v_{i})}{P_{X_{i}|U_{i}V_{i}}(x_{i}|u_{i},v_{i})P_{Y_{i}|U_{i}V_{i}}(y_{i}|u_{i},v_{i})}\leq\eta\Big\},\\
\calB_{3}:= & \Big\{(x^{n},y^{n},u^{n},v^{n}):\nonumber \\
 & \frac{1}{n}\sum_{i=1}^{n}\log\frac{Q_{X_{i}Y_{i}|U_{i}V_{i}}(x_{i},y_{i}|u_{i},v_{i})}{\pi_{XY}(x_{i},y_{i})}\leq R+\eta\Big\}.\label{eq:B3}
\end{align}

We first present a non-asymptotic lower bound on $\left|P_{X^{n}Y^{n}}-\pi_{X^{n}Y^{n}}\right|$,
i.e., a non-asymptotic converse bound for the problem.
\begin{lem}
\label{fblidlossy} For any synthesis code such that
\begin{align}
\frac{1}{n}\log|{\cal M}_{n}| & \leq R,
\end{align}
we have
\begin{equation}
\left|P_{X^{n}Y^{n}}-\pi_{X^{n}Y^{n}}\right|\geq1-P\Big(\bigcap_{i=1}^{3}\calA_{i}\,\Big|\,\widetilde{\calA}\Big)-3\e^{-n\eta},\label{expupppc-2}
\end{equation}
where $P(\cdot|\widetilde{\calA})=P_{X^{n}Y^{n}M_{n}|\widetilde{\calA}}$
denotes the conditional distribution of $(X^{n},Y^{n},M_{n})\sim P_{M_{n}X^{n}Y^{n}}$
given that $(X^{n},Y^{n},M_{n})\in\widetilde{\calA}$, with $P_{M_{n}X^{n}Y^{n}}$
denoting the distribution induced by the synthesis code.
\end{lem}
The proof of Lemma \ref{fblidlossy} is given in Appendix \ref{prooffblidlossy}.

Invoking Lemma \ref{fblidlossy} and choosing the distributions $Q_{X^{n}Y^{n}}$
and $Q_{X^{n}Y^{n}|M_{n}}$ as in the paragraph above \eqref{eq:B1},
we obtain the following lemma.
\begin{lem}
\label{fblidlossy2} Given the conditions in Lemma \ref{fblidlossy},
we have
\begin{equation}
\left|P_{X^{n}Y^{n}}-\pi_{X^{n}Y^{n}}\right|\geq1-P\Big(\bigcap_{i=1}^{3}\calB_{i}\,\Big|\,\widetilde{\calA}\Big)-3\e^{-n\eta}.\label{expupppc-2-1}
\end{equation}
\end{lem}
The proof of Lemma \ref{fblidlossy2} is given in Appendix \ref{prooffblidlossy2}.

In the following, for simplicity, we will use $Q_{i}$ to denote $Q_{X_{i}Y_{i}U_{i}V_{i}}$
and use $P_{i}$ to denote $P_{X_{i}Y_{i}U_{i}V_{i}}$. Let $\alpha\in[0,1]$.
Then we need the following definitions to further lower bound \eqref{expupppc-2-1}.
Similar to the definition of $\omega_{Q_{XYU}}^{(\alpha)}(x,y|u)$
in \eqref{eq:omega}, we define
\begin{align}
 & \omega_{Q_{i},P_{i}}^{(\alpha)}(x_{i},y_{i}|u_{i},v_{i})\nonumber \\*
 & \quad:=\bar{\alpha}\Bigg(\log\frac{Q_{X_{i}Y_{i}|V_{i}}(x_{i},y_{i}|v_{i})}{\pi_{XY}(x_{i},y_{i})}\nonumber \\
 & \qquad+\log\frac{Q_{X_{i}Y_{i}|U_{i}V_{i}}(x_{i},y_{i}|u_{i},v_{i})}{P_{X_{i}|U_{i}V_{i}}(x_{i}|u_{i},v_{i})P_{Y_{i}|U_{i}V_{i}}(y_{i}|u_{i},v_{i})}\Bigg)\nonumber \\
 & \qquad+\alpha\log\frac{Q_{X_{i}Y_{i}|U_{i}V_{i}}(x_{i},y_{i}|u_{i},v_{i})}{\pi_{XY}(x_{i},y_{i})}.\label{eq:omegaQP}
\end{align}
Then, similar to the definition of $\Omega^{(\alpha,\theta)}(Q_{XYU})$
in \eqref{eq:OmegaQ}, we define
\begin{align}
 & \Omega^{(\alpha,\lambda)}(\{Q_{i}\}_{i=1}^{n})\nonumber \\
 &:= -\log\biggl(\sum_{x^{n},y^{n},m}P_{X^{n}Y^{n}M_{n}|\widetilde{\calA}}(x^{n},y^{n},m)\nonumber \\
 & \qquad\times\exp\Big(-\lambda\sum_{i=1}^{n}\omega_{Q_{i},P_{i}}^{(\alpha)}(x_{i},y_{i}|u_{i},v_{i})\Big)\biggr).\label{eq:OmegaQi}
\end{align}
where $u_{i}=m$, $v_{i}=(x^{i-1},y^{i-1})$, and $P_{X^{n}Y^{n}M_{n}|\widetilde{\calA}}$
is the conditional distribution of $(X^{n},Y^{n},M_{n})$ given $(X^{n},Y^{n},M_{n})\in\widetilde{\calA}$.

Applying Cram\'er's bound~\cite[Section~2.2]{Dembo} and utilizing
Lemma \ref{fblidlossy2}, we obtain the following lemma. The proof
of this lemma is similar to that of~\cite[Proposition~1]{oohama2016exponent},
and hence we omit it for the sake of brevity.
\begin{lem}
\label{fblidlossy3} For any $(\alpha,\lambda)\in[0,1]\times[0,\infty)$,
given the condition in Lemma \ref{fblidlossy}, we have
\begin{align}
 & \left|P_{X^{n}Y^{n}}-\pi_{X^{n}Y^{n}}\right|\nonumber \\
 & \geq1-4\exp\left(-n\frac{\frac{1}{n}\Omega^{(\alpha,\lambda)}(\{Q_{i}\}_{i=1}^{n})-\lambda\alpha R}{1+(1+\bar{\alpha})\lambda}\right).
\end{align}
\end{lem}
Let
\begin{align}
\underline{\Omega}^{(\alpha,\lambda)}:=\inf_{n\geq1}\inf_{\{Q_{i}\}_{i=1}^{n}}\frac{1}{n}\Omega^{(\alpha,\lambda)}(\{Q_{i}\}_{i=1}^{n}).\label{def:lossyunderlineOmega}
\end{align}
Define
\begin{align}
\theta:=\frac{\lambda}{1-2\bar{\alpha}\lambda}.\label{def:theta}
\end{align}
Hence, we have
\begin{align}
\lambda=\frac{\theta}{1+2\bar{\alpha}\theta}.\label{lambdausetheta}
\end{align}

The next lemma is essential in the proof.
\begin{lem}
\label{lbunderlineOmegalossy} For $\alpha\in[0,1]$ and $\lambda\in[0,\frac{1}{2\bar{\alpha}})$,
we have
\begin{align}
\underline{\Omega}^{(\alpha,\lambda)}\geq\frac{\Omega^{(\alpha,\theta)}}{1+2\bar{\alpha}\theta}.
\end{align}
\end{lem}
The proof of Lemma \ref{lbunderlineOmegalossy} is similar to that
of \cite[Proposition~2]{oohama2016exponent} and given in Appendix
\ref{prooflbunderlineOmegalossy}. In the proof of Lemma \ref{lbunderlineOmegalossy},
we adopt ideas from \cite{oohama2016exponent} and choose appropriate
distributions $Q_{X_{i}Y_{i}U_{i}V_{i}}$ via the recursive method.

Combining Lemmas \ref{fblidlossy3} and \ref{lbunderlineOmegalossy}
yields
\begin{align}
 & \left|P_{X^{n}Y^{n}}-\pi_{X^{n}Y^{n}}\right|\nonumber \\
 & \geq1-4\exp\left(-n\frac{\underline{\Omega}^{(\alpha,\lambda)}-\lambda\alpha R}{1+(1+\bar{\alpha})\lambda}\right)\\
 & \geq1-4\exp\left(-n\frac{\frac{\Omega^{(\alpha,\theta)}}{1+2\bar{\alpha}\theta}-\frac{\theta\alpha R}{1+2\bar{\alpha}\theta}}{1+\frac{(1+\bar{\alpha})\theta}{1+2\bar{\alpha}\theta}}\right)\\
 & =1-4\exp\Bigg(-n\frac{\Omega^{(\alpha,\theta)}-\theta\alpha R}{1+(5-3\alpha)\theta}\Bigg)\label{eq:-37}\\
 & \geq1-4\exp\big(-nF(R)\big),\label{useflossy}
\end{align}
where \eqref{useflossy} follows from the definition of $F(R)$ in
\eqref{def:lossyF} and the fact that \eqref{eq:-37} holds for any
$(\alpha,\theta)\in[0,1]\times(0,+\infty)$. The proof of Theorem
\ref{thm:expstrongconverseforTV0} is now complete.

\subsection{Proof of Lemma \ref{fblidlossy}}

\label{prooffblidlossy}

Define $\pi_{X^{n}Y^{n}M_{n}}:=\pi_{X^{n}Y^{n}}P_{M_{n}|X^{n}Y^{n}}$.
Then
\begin{align}
 & \left|P_{X^{n}Y^{n}}-\pi_{X^{n}Y^{n}}\right|\nonumber \\
 & =\left|P_{X^{n}Y^{n}M_{n}}-\pi_{X^{n}Y^{n}M_{n}}\right|\allowdisplaybreaks\\
 & \geq\pi(\widetilde{\calA}_{1}\cap\calA_{1}\cap\calA_{3})-P(\widetilde{\calA}_{1}\cap\calA_{1}\cap\calA_{3})\allowdisplaybreaks\\
 & =1-\pi(\widetilde{\calA}_{1}^{c}\cup\calA_{1}^{c}\cup\calA_{3}^{c})-P(\widetilde{\calA}_{1}\cap\calA_{1}\cap\calA_{3})\allowdisplaybreaks\\
 & =1-P\Big(\widetilde{\calA}\cap\Big(\bigcap_{i=1}^{3}\calA_{i}\Big)\Big)\nonumber \\
 & \qquad-P(\widetilde{\calA}_{1}\cap\calA_{1}\cap\calA_{3}\cap(\calA_{2}^{c}\cup\widetilde{\calA}_{2}^{c}))\allowdisplaybreaks\nonumber \\
 & \qquad-\pi(\widetilde{\calA}_{1}^{c}\cup\calA_{1}^{c}\cup\calA_{3}^{c})\allowdisplaybreaks\\
 & \geq1-P\Big(\widetilde{\calA}\cap\Big(\bigcap_{i=1}^{3}\calA_{i}\Big)\Big)-P(\calA_{2}^{c})-P(\widetilde{\calA}_{2}^{c})\nonumber \\
 & \qquad-\pi(\widetilde{\calA}_{1}^{c})-\pi(\calA_{1}^{c})-\pi(\calA_{3}^{c})\\
 & =1-P\Big(\widetilde{\calA}\cap\Big(\bigcap_{i=1}^{3}\calA_{i}\Big)\Big)-P(\calA_{2}^{c})-\pi(\calA_{1}^{c})-\pi(\calA_{3}^{c})
\end{align}
The last three terms above can each be bounded above by $\e^{-n\eta}$
because
\begin{align}
P(\calA_{2}^{c}) & =\sum_{\substack{(x^{n},y^{n},m)\in\calA_{2}^{c}}
}P(x^{n},y^{n},m)\\
 & \leq\sum_{\substack{(x^{n},y^{n},m)\in\calA_{2}^{c}}
}P(w)Q(x^{n},y^{n}|m)\e^{-n\eta}\label{ignorecons}\\
 & \leq\e^{-n\eta},\label{uppDelta2pcnidlossy}
\end{align}
and
\begin{align}
\pi(\calA_{3}^{c}) & =\sum_{\substack{(x^{n},y^{n},m)\in\calA_{3}^{c}}
}\pi(x^{n},y^{n})P(m|x^{n},y^{n})\label{ignorecons-2}\\
 & \leq\sum_{\substack{(x^{n},y^{n},m)\in\calA_{3}^{c}}
}Q(x^{n},y^{n}|m)\nonumber \\
 & \qquad\qquad\times\e^{-n(R+\eta)}P(m|x^{n},y^{n})\\
 & \leq\sum_{\substack{(x^{n},y^{n},m)\in\calA_{3}^{c}}
}Q(x^{n},y^{n}|m)\e^{-n(R+\eta)}\\
 & \leq\e^{-n\eta},\label{uppDelta2pcnidlossy-2}
\end{align}
and
\begin{align}
\pi(\calA_{1}^{c}) & =\sum_{\substack{(x^{n},y^{n})\in\calA_{1}^{c}}
}\pi(x^{n},y^{n})\\
 & \leq\sum_{\substack{(x^{n},y^{n})\in\calA_{1}^{c}}
}Q(x^{n},y^{n})\e^{-n\eta}\label{ignorecons-3}\\
 & \leq\e^{-n\eta}.\label{uppDelta2pcnidlossy-3}
\end{align}
Therefore, we have
\begin{align}
 & \left|P_{X^{n}Y^{n}}-\pi_{X^{n}Y^{n}}\right|\nonumber \\
 & \geq1-P\Big(\widetilde{\calA}\cap\Big(\bigcap_{i=1}^{3}\calA_{i}\Big)\Big)-3\e^{-n\eta}\\
 & \geq1-P\Big(\bigcap_{i=1}^{3}\calA_{i}\,\Big|\,\widetilde{\calA}\Big)-3\e^{-n\eta}.
\end{align}

\subsection{Proof of Lemma \ref{fblidlossy2}}

\label{prooffblidlossy2}

Recall that in Appendix \ref{subsec:Proof-of-Theoremexpstrong}, we
choose $U_{i}=M_{n}$ and $V_{i}=(X^{i-1},Y^{i-1})$. Then $Q_{X^{n}Y^{n}}$
and $Q_{X^{n}Y^{n}|M_{n}}$ can be written as follows:
\begin{align}
 & Q_{X^{n}Y^{n}}(x^{n},y^{n})\nonumber \\
 & =\prod_{i=1}^{n}Q_{X_{i}Y_{i}|X^{i-1}Y^{i-1}}(x_{i},y_{i}|x^{i-1},y^{i-1})\label{eq:-21}\\
 & =\prod_{i=1}^{n}Q_{X_{i}Y_{i}|V_{i}}(x_{i},y_{i}|v_{i}),\allowdisplaybreaks\\
 & Q_{X^{n}Y^{n}|M_{n}}(x^{n},y^{n}|m)\allowdisplaybreaks\nonumber \\*
 & =\prod_{i=1}^{n}Q_{X_{i}Y_{i}|M_{n}X^{i-1}Y^{i-1}}(x_{i},y_{i}|m,x^{i-1},y^{i-1})\allowdisplaybreaks\\
 & =\prod_{i=1}^{n}Q_{X_{i}Y_{i}|U_{i}V_{i}}(x_{i},y_{i}|u_{i},v_{i}).
\end{align}
Now recall from Appendix \ref{subsec:Proof-of-Theoremexpstrong} that
the joint distribution of $(X^{n},Y^{n},M_{n})$ induced by the code
is $P_{X^{n}Y^{n}M_{n}}$. The marginal distributions of $P_{X^{n}Y^{n}M_{n}}$
are as follows:
\begin{align}
\pi_{X^{n}Y^{n}}(x^{n},y^{n}) & =\prod_{i=1}^{n}\pi_{XY}(x_{i},y_{i}),\label{lossypy}\\
P_{X^{n}|M_{n}}(x^{n}|m) & =\prod_{i=1}^{n}P_{X_{i}|M_{n}X^{i-1}}(x_{i}|m,x^{i-1})\allowdisplaybreaks\\
 & =\prod_{i=1}^{n}P_{X_{i}|M_{n}X^{i-1}Y^{i-1}}(x_{i}|m,x^{i-1},y^{i-1})\allowdisplaybreaks\label{eq:-30}\\
 & =\prod_{i=1}^{n}P_{X_{i}|U_{i}V_{i}}(x_{i}|u_{i},v_{i}),\allowdisplaybreaks\\
P_{Y^{n}|M_{n}}(y^{n}|m) & =\prod_{i=1}^{n}P_{Y_{i}|M_{n}Y^{i-1}}(y_{i}|m,y^{i-1})\allowdisplaybreaks\\
 & =\prod_{i=1}^{n}P_{Y_{i}|M_{n}X^{i-1}Y^{i-1}}(y_{i}|m,x^{i-1},y^{i-1})\label{eq:-31}\\
 & =\prod_{i=1}^{n}P_{Y_{i}|U_{i}V_{i}}(y_{i}|u_{i},v_{i}),\label{eq:-22}
\end{align}
where \eqref{eq:-30} and \eqref{eq:-31} follow from the Markov chains
$X_{i}-M_{n}X^{i-1}-Y^{i-1}$ and $Y_{i}-M_{n}Y^{i-1}-X^{i-1}$ under
distribution $P_{X^{n}Y^{n}M_{n}}$ (these two Markov chains can be
easily obtained by observing that $P_{X^{i}Y^{i}M_{n}}=P_{M_{n}}P_{X_{i}|M_{n}X^{i-1}}P_{Y_{i}|M_{n}Y^{i-1}}$).

Using Lemma \ref{fblidlossy} and \eqref{eq:-21}\textendash \eqref{eq:-22},
we obtain
\begin{equation}
\left|P_{X^{n}Y^{n}}-\pi_{X^{n}Y^{n}}\right|\geq1-P\Big(\bigcap_{i=1}^{3}\calB_{i}\,\Big|\,\widetilde{\calA}\Big)-3\e^{-n\eta}.\label{expupppc-2-1-1}
\end{equation}

\subsection{Proof of Lemma \ref{lbunderlineOmegalossy}}

\label{prooflbunderlineOmegalossy}

\subsubsection{Removing Dependence on the Indices}

Recall from Appendix \ref{subsec:Proof-of-Theoremexpstrong} that
the joint distribution of $(X^{n},Y^{n},M_{n})$ is $P_{X^{n}Y^{n}M_{n}}$
and $P_{X_{i}Y_{i}U_{i}V_{i}}$ is induced by $P_{X^{n}Y^{n}M_{n}}$.
Further, recall that $Q_{i}$ denotes $Q_{X_{i}Y_{i}U_{i}V_{i}}$
and $P_{i}$ denotes $P_{X_{i}Y_{i}U_{i}V_{i}}$. Define
\begin{align}
g_{Q_{i},P_{i}}^{(\alpha,\lambda)}(x_{i},y_{i}|u_{i},v_{i}) & :=\exp\Big(-\lambda\omega_{Q_{i},P_{i}}^{(\alpha)}(x_{i},y_{i}|u_{i},v_{i})\Big),\label{flambda}
\end{align}
where $\omega_{Q_{i},P_{i}}^{(\alpha)}(x_{i},y_{i}|u_{i},v_{i})$
is defined in \eqref{eq:omegaQP}.

Recall the definition of $\Omega^{(\alpha,\lambda)}(\{Q_{i}\}_{i=1}^{n})$
in \eqref{eq:OmegaQi}, then we obtain that
\begin{align}
 & \exp\Big(-\Omega^{(\alpha,\lambda)}(\{Q_{i}\}_{i=1}^{n})\Big)\nonumber \\
 & =\sum_{x^{n},y^{n},m}P_{X^{n}Y^{n}M_{n}|\widetilde{\calA}}(x^{n},y^{n},m)\prod_{i=1}^{n}g_{Q_{i},P_{i}}^{(\alpha,\lambda)}(x_{i},y_{i}|u_{i},v_{i}),\label{lossyOmeganow}
\end{align}
where $u_{i}=m$, $v_{i}=(x^{i-1},y^{i-1})$, and $P_{X^{n}Y^{n}M_{n}|\widetilde{\calA}}$
is the conditional distribution of $(X^{n},Y^{n},M_{n})$ given $(X^{n},Y^{n},M_{n})\in\widetilde{\calA}$.

For $i=1,\ldots,n$, define
\begin{align}
 & \tilC_{i}:=\sum_{x^{n},y^{n},m}P_{X^{n}Y^{n}M_{n}|\widetilde{\calA}}(x^{n},y^{n},m)\nonumber \\
 & \qquad\times\prod_{j=1}^{i}g_{Q_{j},P_{j}}^{(\alpha,\lambda)}(x_{j},y_{j}|u_{j},v_{j}),\label{def:tilCi}\\
 & P_{X^{n}Y^{n}M_{n}|\widetilde{\calA}}^{(\alpha,\lambda)|i}(x^{n},y^{n},m):=\frac{1}{\tilC_{i}}P_{X^{n}Y^{n}M_{n}|\widetilde{\calA}}(x^{n},y^{n},m)\nonumber \\
 & \qquad\times\prod_{j=1}^{i}g_{Q_{j},P_{j}}^{(\alpha,\lambda)}(x_{j},y_{j}|u_{j},v_{j}),\label{def:PSZiambl}\\
 & \Lambda_{i}^{(\alpha,\lambda)}(\{Q_{j}\}_{j=1}^{i}):=\frac{\tilC_{i}}{\tilC_{i-1}}.\label{def:Lambdai}
\end{align}
Obviously, $P_{X^{n}Y^{n}M_{n}|\widetilde{\calA}}^{(\alpha,\lambda)|i}(x^{n},y^{n},m)$
is a distribution induced by normalizing all the terms of the summation
in the definition of $\tilC_{i}$.

Similarly to \cite[Lemma 7]{oohama2016exponent}, we obtain the following
lemma, which will be used to simplify $\Lambda_{i}^{(\alpha,\lambda)}(\{Q_{j}\}_{j=1}^{i})$,
defined in \eqref{def:Lambdai}, in Appendix \ref{subsubsec:Completion}.

\begin{lem}
\label{lem:Lambdai} For $i=1,\ldots,n$, we have
\begin{align}
 & \Lambda_{i}^{(\alpha,\lambda)}(\{Q_{j}\}_{j=1}^{i})\nonumber \\
 & =\sum_{x^{n},y^{n},m}P_{X^{n}Y^{n}M_{n}|\widetilde{\calA}}^{(\alpha,\lambda)|i-1}(x^{n},y^{n},m)g_{Q_{i},P_{i}}^{(\alpha,\lambda)}(x_{i},y_{i}|u_{i},v_{i}).
\end{align}
\end{lem}
Furthermore, combining \eqref{lossyOmeganow}, \eqref{def:tilCi}
and \eqref{def:Lambdai} gives us
\begin{align}
\exp\Big(-\Omega^{(\alpha,\lambda)}(\{Q_{i}\}_{i=1}^{n})\Big)=\prod_{i=1}^{n}\Lambda_{i}^{(\alpha,\lambda)}(\{Q_{j}\}_{j=1}^{i}).\label{lossyOmegaow4}
\end{align}

\subsubsection{Completion of the Proof of Lemma \ref{lbunderlineOmegalossy}}

\label{subsubsec:Completion} Assume $\mathcal{U}$ and $\mathcal{V}$
are two countable sets. Paralleling~\eqref{eq:Q} to~\eqref{eq:Omega},
for $(\alpha,\theta)\in(0,1]\times(0,\infty)$, we define the following
quantities:
\begin{align}
 & \widetilde{\calQ}:=\Big\{ Q_{XYUV}\in\calP(\calX\times\calY\times\calU\times\calV):\nonumber \\*
 & \qquad~\mathrm{supp}(Q_{XY})\subseteq\mathrm{supp}(\pi_{XY})\Big\},\label{eq:OmegaQ-2}\\
 & \widetilde{\omega}_{Q_{XYUV}}^{(\alpha)}(x,y|u,v):=\bar{\alpha}\Bigg(\log\frac{Q_{XY|V}(x,y|v)}{\pi_{XY}(x,y)}\nonumber \\
 & \qquad+\log\frac{Q_{XY|UV}(x,y|u,v)}{Q_{X|UV}(x|u,v)Q_{Y|UV}(y|u,v)}\Bigg)\nonumber \\
 & \qquad+\alpha\log\frac{Q_{XY|UV}(x,y|u,v)}{\pi_{XY}(x,y)},\\
 & \widetilde{\Omega}^{(\alpha,\lambda)}(Q_{XYUV})\nonumber \\
 & \quad:=-\log\mathbb{E}_{Q_{XYUV}}\Big[\exp\big(-\theta\omega_{Q_{XYUV}}^{(\alpha)}(X,Y|U,V)\big)\Big],\label{eq:-17-1}\\
 & \widetilde{\Omega}^{(\alpha,\lambda)}:=\inf_{Q_{XYUV}\in\widetilde{\calQ}}\widetilde{\Omega}^{(\alpha,\lambda)}(Q_{XYUV}),\label{eq:Omegaoverline}
\end{align}
where $\mathbb{E}_{Q_{XYUV}}$ in \eqref{eq:-17-1} is only taken
over the set $\mathrm{supp}(Q_{XYUV})$.

Recall that $u_{i}=m$ and $v_{i}=(x^{i-1},y^{i-1})$. For each $i=1,\ldots,n$,
define
\begin{align}
P^{(\alpha,\lambda)}(x_{i},y_{i},u_{i},v_{i}) & :=\sum_{x_{i+1}^{n},y_{i+1}^{n}}P_{X^{n}Y^{n}M_{n}|\widetilde{\calA}}^{(\alpha,\lambda)|i-1}(x^{n},y^{n},m),\label{def:PSXYZhXambl}
\end{align}
where $P_{X^{n}Y^{n}M_{n}}^{(\alpha,\lambda)|i-1}(x^{n},y^{n},m)$
was defined in \eqref{def:PSZiambl}.

Combining Lemma \ref{lem:Lambdai} and \eqref{def:PSXYZhXambl} yields
\begin{align}
 & \Lambda_{i}^{(\alpha,\lambda)}(\{Q_{j}\}_{j=1}^{i})\nonumber \\
 & =\sum_{x_{i},y_{i},u_{i},v_{i}}P^{(\alpha,\lambda)}(x_{i},y_{i},u_{i},v_{i})g_{Q_{i},P_{i}}^{(\alpha,\lambda)}(x_{i},y_{i}|u_{i},v_{i}).\label{Lambdaieq}
\end{align}

Note that $Q_{i}=Q_{X_{i}Y_{i}U_{i}V_{i}}$ can be chosen arbitrarily
for all $i=1,\ldots,n$. Here we apply the recursive method. For each
$i=1,\ldots,n$, we choose $Q_{X_{i}Y_{i}U_{i}V_{i}}$ such that
\begin{align}
Q_{X_{i}Y_{i}U_{i}V_{i}}(x_{i},y_{i},u_{i},v_{i}) & =P^{(\alpha,\lambda)}(x_{i},y_{i},u_{i},v_{i}).\label{eq:Qi}
\end{align}
Then, let $Q_{X_{i}Y_{i}|V_{i}},Q_{X_{i}Y_{i}|U_{i}V_{i}}$ be induced
by $Q_{X_{i}Y_{i}U_{i}V_{i}}$.

Define
\begin{align}
 & h_{Q_{i}}^{(\alpha,\lambda)}(x_{i},y_{i}|u_{i},v_{i}):=g_{Q_{i},P_{i}}^{(\alpha,\lambda)}(x_{i},y_{i}|u_{i},v_{i})\nonumber \\
 & \qquad\times\Bigg(\frac{P_{X_{i}|U_{i}V_{i}}^{\lambda\bar{\alpha}}(x_{i}|u_{i},v_{i})P_{Y_{i}|U_{i}V_{i}}^{\lambda\bar{\alpha}}(y_{i}|u_{i},v_{i})}{Q_{X_{i}|U_{i}V_{i}}^{\lambda\bar{\alpha}}(x_{i}|u_{i},v_{i})Q_{Y_{i}|U_{i}V_{i}}^{\lambda\bar{\alpha}}(y_{i}|u_{i},v_{i})}\Bigg)^{-1},\label{eq:hQi}
\end{align}
where $g_{Q_{i},P_{i}}^{(\alpha,\lambda)}$ was defined in \eqref{flambda}.
In the following, for brevity, we drop the subscripts of the distributions.
From~\eqref{Lambdaieq}, we obtain
\begin{align}
 & \Lambda_{i}^{(\alpha,\lambda)}(\{Q_{j}\}_{j=1}^{i})\nonumber \\
 & =\mathbb{E}_{Q_{i}}[g_{Q_{i},P_{i}}^{(\alpha,\lambda)}(X_{i},Y_{i}|U_{i},V_{i})]\\
 & =\mathbb{E}_{Q_{i}}\Bigg[h_{Q_{i}}^{(\alpha,\lambda)}(X_{i},Y_{i}|U_{i},V_{i})\nonumber \\
 & \qquad\times\frac{P_{X_{i}|U_{i}V_{i}}^{\lambda\bar{\alpha}}(X_{i}|U_{i},V_{i})P_{Y_{i}|U_{i}V_{i}}^{\lambda\bar{\alpha}}(Y_{i}|U_{i},V_{i})}{Q_{X_{i}|U_{i}V_{i}}^{\lambda\bar{\alpha}}(X_{i}|U_{i},V_{i})Q_{Y_{i}|U_{i}V_{i}}^{\lambda\bar{\alpha}}(Y_{i}|U_{i},V_{i})}\Bigg]\label{usefcompare}\\
 & \leq\Big(\mathbb{E}_{Q_{i}}\Big[\Big\{ h_{Q_{i}}^{(\alpha,\lambda)}(X_{i},Y_{i}|U_{i},V_{i})\Big\}^{\frac{1}{1-2\lambda\bar{\alpha}}}\Big]\Big)^{1-2\lambda\bar{\alpha}}\nonumber \\
 & \qquad\times\Bigg(\mathbb{E}_{Q_{i}}\Bigg[\frac{P_{X_{i}|U_{i}V_{i}}(X_{i}|U_{i},V_{i})}{Q_{X_{i}|U_{i}V_{i}}(X_{i}|U_{i},V_{i})}\Bigg]\Bigg)^{\lambda\bar{\alpha}}\nonumber \\
 & \qquad\times\Bigg(\mathbb{E}_{Q_{i}}\Bigg[\frac{P_{Y_{i}|U_{i}V_{i}}(Y_{i}|U_{i},V_{i})}{Q_{Y_{i}|U_{i}V_{i}}(Y_{i}|U_{i},V_{i})}\Bigg]\Bigg)^{\lambda\bar{\alpha}}\label{useholderineq}\\
 & \leq\exp\Big(-\big(1-2\lambda\bar{\alpha}\big)\widetilde{\Omega}^{(\alpha,\frac{\lambda}{1-2\lambda\bar{\alpha}})}(Q_{i})\Big)\label{usedeflossyOmegaambl}\\
 & =\exp\left(-\frac{\widetilde{\Omega}^{(\alpha,\theta)}(Q_{i})}{1+2\bar{\alpha}\theta}\right)\label{usetheta}\\
 & \leq\exp\left(-\frac{\widetilde{\Omega}^{(\alpha,\theta)}}{1+2\bar{\alpha}\theta}\right)\label{usehOmeganambl}\\*
 & =\exp\left(-\frac{\Omega^{(\alpha,\theta)}}{1+2\bar{\alpha}\theta}\right),\label{uselemmacard}
\end{align}
where \eqref{usefcompare} follows from \eqref{eq:hQi}; \eqref{useholderineq}
follows from H\"older's inequality; \eqref{usedeflossyOmegaambl} follows
from the definitions of $\Omega^{(\alpha,\theta)}(\cdot)$ and $h_{Q_{i}}^{(\alpha,\lambda)}(\cdot)$
in \eqref{eq:OmegaQ} and \eqref{eq:hQi} respectively; \eqref{usetheta}
follows from \eqref{def:theta} and \eqref{lambdausetheta}; \eqref{usehOmeganambl}
follows since $\Omega^{(\alpha,\theta)}(Q_{XYUV})\geq\widetilde{\Omega}^{(\alpha,\theta)}$
for any $Q_{XYUV}$ such that $\mathrm{supp}(Q_{XY})\subseteq\mathrm{supp}(\pi_{XY})$
(The fact that $Q_{i}$ satisfies this point will be shown in the
following paragraph); and \eqref{uselemmacard} follows since by the
support lemma \cite{Gamal}, the cardinality bounds $|\mathcal{V}|\leq1$,
$|\calU|\leq|\calX||\calY|$ are sufficient to exhaust~$\widetilde{\Omega}^{(\alpha,\theta)}$.

Now we show that according to the choice of $Q_{X_{i}Y_{i}U_{i}V_{i}}$,
we have $\mathrm{supp}(Q_{X_{i}Y_{i}})\subseteq\mathrm{supp}(\pi_{XY})$,
which was used in \eqref{usehOmeganambl}. Note that $P_{X^{n}Y^{n}M_{n}}(x^{n},y^{n},m)>0$
and $\pi_{X^{n}Y^{n}}(x^{n},y^{n})>0$ for any $(x^{n},y^{n},m)\in\widetilde{\calA}$,
and hence the marginal distributions $P_{X_{i}Y_{i}U_{i}V_{i}}$,
$P_{X_{i}|U_{i}V_{i}}$ and $P_{Y_{i}|U_{i}V_{i}}$ when evaluated
at any $(x^{n},y^{n},m)\in\widetilde{\calA}$ is positive as well.
According to the choice of $Q_{X_{i}Y_{i}U_{i}V_{i}}$ in \eqref{eq:Qi},
we have that $Q_{X_{i}Y_{i}U_{i}V_{i}}$ is also positive when evaluated
at $(x^{n},y^{n},m)\in\widetilde{\calA}$ (this point can be shown
via mathematical induction), i.e.,
\begin{align}
 & \mathrm{supp}(Q_{X_{i}Y_{i}U_{i}V_{i}})\supseteq\bigl\{(x,y,u,v):\exists(x^{n},y^{n},m)\in\widetilde{\calA}:\nonumber \\
 & \qquad\qquad\qquad\qquad x_{i}=x,y_{i}=y,m=u,(x^{i-1},y^{i-1})=v\bigr\}.
\end{align}
On the other hand, also according to the choice of $Q_{X_{i}Y_{i}U_{i}V_{i}}$,
we have
\begin{align}
 & \mathrm{supp}(Q_{X_{i}Y_{i}U_{i}V_{i}})\subseteq\bigl\{(x,y,u,v):\exists(x^{n},y^{n},m)\in\widetilde{\calA}:\nonumber \\
 & \qquad\qquad\qquad\qquad x_{i}=x,y_{i}=y,m=u,(x^{i-1},y^{i-1})=v\bigr\}.
\end{align}
Therefore,
\begin{align}
 & \mathrm{supp}(Q_{X_{i}Y_{i}U_{i}V_{i}})=\bigl\{(x,y,u,v):\exists(x^{n},y^{n},m)\in\widetilde{\calA}:\nonumber \\
 & \qquad\qquad\qquad\qquad x_{i}=x,y_{i}=y,m=u,(x^{i-1},y^{i-1})=v\bigr\}.
\end{align}
Further, we have
\begin{align}
 & \mathrm{supp}(Q_{X_{i}Y_{i}})\nonumber \\
 & =\left\{ (x,y):\exists(x^{n},y^{n},m)\in\widetilde{\calA}:x_{i}=x,y_{i}=y\right\} \\
 & \subseteq\bigl\{(x,y):\exists(x^{n},y^{n},m)\in\mathrm{supp}(\pi_{X^{n}Y^{n}})\times{\cal M}_{n}:\nonumber \\
 & \qquad\qquad x_{i}=x,y_{i}=y\bigr\}\\
 & =\mathrm{supp}(\pi_{XY}).\label{eq:-18}
\end{align}

Combining \eqref{lossyOmegaow4} and \eqref{uselemmacard}, we obtain
that
\begin{align}
\frac{1}{n}\Omega^{(\alpha,\lambda)}(\{Q_{i}\}_{i=1}^{n}) & =-\frac{1}{n}\sum_{i=1}^{n}\log\Lambda_{i}^{(\alpha,\lambda)}(\{Q_{j}\}_{j=1}^{i})\\*
 & \geq\frac{\Omega^{(\alpha,\theta)}}{1+2\bar{\alpha}\theta}.\label{finallyready}
\end{align}
Finally, combining~\eqref{def:lossyunderlineOmega} and \eqref{finallyready},
we have that
\begin{align}
\underline{\Omega}^{(\alpha,\lambda)} & \geq\frac{\Omega^{(\alpha,\theta)}}{1+2\bar{\alpha}\theta}.
\end{align}
The proof of Lemma \ref{lbunderlineOmegalossy} is now complete.

\section{Proof of Lemma \ref{propFOmega}}

\label{proofpropFOmega}

Let $U$ be a random variable taking values in a finite alphabet $\calU$.
Define a set of joint distributions on $\calX\times\calY\times\calU$
as
\begin{align}
\calP^{*} & :=\Big\{ P_{XYU}:~|\calU|\leq|\calX||\calY|,~P_{XY}=\pi_{XY},~X-U-Y\Big\}.
\end{align}
and let
\begin{align}
R^{*} & :=\min_{P_{XYU}\in\calP^{*}}I(XY;U).\label{eq:Rstar}
\end{align}

\subsection{Preliminary Lemmata for the Proof of Lemma \ref{propFOmega}}

By the support lemma \cite[Appendix~C]{Gamal}, we have the following
lemma \cite{Wyner}.
\begin{lem}
\label{lem:wyner} Wyner's common information $C_{\mathsf{Wyner}}(X;Y)$
satisfies
\begin{align}
C_{\mathsf{Wyner}}(X;Y) & =R^{*}.
\end{align}
\end{lem}
Before proceeding the proof of Lemma \ref{propFOmega}, we present
an alternative expression for Wyner's common information. Recall that
given a number $a\in[0,1]$, we define $\bar{a}=1-a$. Then for any
$\alpha\in[0,1]$ and $Q_{XYU}\in\calQ$, define
\begin{align}
R^{(\alpha)}(Q_{XYU}) & :=\bar{\alpha}\big(D(Q_{XY}\|\pi_{XY})\nonumber \\
 & \qquad+D(Q_{XY|U}\|Q_{X|U}Q_{Y|U}|Q_{U})\big)\nonumber \\
 & \qquad+\alpha D(Q_{XY|U}\|\pi_{XY}|Q_{U}),\label{eq:RaQ}\\
R^{(\alpha)} & :=\min_{Q_{XYU}\in\calQ}R^{(\alpha)}(Q_{XYU})\label{eq:Ra}\\
R_{\mathrm{sh}} & :=\sup_{\alpha\in(0,1]}\frac{1}{\alpha}R^{(\alpha)}.\label{eq:Rsh}
\end{align}
By observing that both $\calP^{*}$ and $\calQ$ are compact, and
by utilizing the fact that a continuous function defined on a compact
set attains its minimum, we obtain the following.
\begin{fact}
\label{fact:finite} Both the minima in the definitions of $R^{*}$
in \eqref{eq:Rstar} and $R^{(\alpha)}$ in \eqref{eq:Ra} are attained.
\end{fact}
We then have the following lemma.
\begin{lem}
\label{propcalRshlossy} The following conclusions hold.
\begin{itemize}
\item[(i)] For any $\alpha\in(0,1]$, we have
\begin{equation}
\frac{1}{\alpha}R^{(\alpha)}\leq R^{*}.\label{eq:Ra-1}
\end{equation}
Moreover, there exists some decreasing sequence $\left\{ \alpha_{k}\right\} _{k=1}^{\infty}\subset\bbR$
such that $\lim_{k\to\infty}\alpha_{k}=0$ and
\begin{equation}
\frac{1}{\alpha_{k}}R^{(\alpha_{k})}\geq R^{*}-c(\alpha_{k}),\label{eq:Ra-2}
\end{equation}
where $\{c(\alpha_{k})\}_{k=1}^{\infty}\subset\bbR$ is another sequence
such that $\lim_{k\to\infty}c(\alpha_{k})=0$.
\item[(ii)] We have
\begin{align}
R_{\mathrm{sh}}=R^{*}=C_{\mathsf{Wyner}}(X;Y).\label{tilcalReqcalrshlossy}
\end{align}
\end{itemize}
\end{lem}
Lemma \ref{propcalRshlossy} is similar to \cite[Property 3]{oohama2016exponent},
but the proofs are different. Essentially, in both the proof of \cite[Property 3]{oohama2016exponent}
and our proof, an intermediate distribution $\widetilde{Q}_{XYU}$
is used to establish the inequality
\begin{equation}
R^{*}-c(\alpha_{k})\leq\frac{1}{\alpha_{k}}R^{(\alpha_{k})}(\widetilde{Q}_{XYU})\leq\frac{1}{\alpha_{k}}R^{(\alpha_{k})}.
\end{equation}
However, the construction of such an intermediate distribution is
different for these two proofs. The construction in~\cite{oohama2016exponent}
does not apply to our case, since our case does not only require $\widetilde{Q}_{XYU}$
to satisfy the Markov chain $X-U-Y$, but also requires that $\widetilde{Q}_{XY}=\pi_{XY}$.
\begin{IEEEproof}[Proof of Lemma \ref{propcalRshlossy}]
It is easy to show \eqref{eq:Ra-1}. Hence, by the definition of
$R_{\mathrm{sh}}$ in \eqref{eq:Rsh},
\begin{equation}
R_{\mathrm{sh}}\leq R^{*}.\label{eq:-8}
\end{equation}
In the following we prove \eqref{eq:Ra-2}. Let $\left\{ \alpha_{m}\right\} _{m=1}^{\infty}$
be an arbitrary sequence of decreasing positive real numbers such
that $\lim_{m\to\infty}\alpha_{m}=0$, and let $Q_{XYU}^{(m)}$ be
a minimizing distribution of \eqref{eq:Ra} with $\alpha=\alpha_{m}$.
The existence of this minimizing distribution is guaranteed by Fact~\ref{fact:finite}.
Since $\calP(\calX\times\calY\times\calU)$ is compact (following
from the definition of $\calQ$), there must exist some sequence of
increasing integers $\left\{ m_{k}\right\} _{k=1}^{\infty}$ such
that $Q_{XYU}^{(m_{k})}$ converges to some distribution $\widetilde{Q}_{XYU}$.
Consider,
\begin{align}
R_{\mathrm{sh}} & =\sup_{\alpha\in(0,1]}\frac{1}{\alpha}R^{(\alpha)}\label{eq:Rsh-1}\\
 & \geq\limsup_{k\to\infty}\frac{1}{\alpha_{m_{k}}}R^{(\alpha_{m_{k}})}\\
 & =\limsup_{k\to\infty}\Bigg\{\frac{\bar{\alpha}_{m_{k}}}{\alpha_{m_{k}}}\Big(D(Q_{XY}^{(m_{k})}\|\pi_{XY})\nonumber \\
 & \qquad+D(Q_{XY|U}^{(m_{k})}\|Q_{X|U}^{(m_{k})}Q_{Y|U}^{(m_{k})}|Q_{U}^{(m_{k})})\Big)\nonumber \\
 & \qquad+D(Q_{XY|U}^{(m_{k})}\|\pi_{XY}|Q_{U}^{(m_{k})})\Bigg\}\\
 & \geq\limsup_{k\to\infty}\left\{ \frac{\bar{\alpha}_{m_{k}}}{\alpha_{m_{k}}}\right\} \liminf_{k\to\infty}\bigl\{ D(Q_{XY}^{(m_{k})}\|\pi_{XY})\nonumber \\
 & \qquad+D(Q_{XY|U}^{(m_{k})}\|Q_{X|U}^{(m_{k})}Q_{Y|U}^{(m_{k})}|Q_{U}^{(m_{k})})\bigr\}\nonumber \\
 & \qquad+\liminf_{k\to\infty}D(Q_{XY|U}^{(m_{k})}\|\pi_{XY}|Q_{U}^{(m_{k})})\\
 & =\infty\big(D(\widetilde{Q}_{XY}\|\pi_{XY})+D(\widetilde{Q}_{XY|U}\|\widetilde{Q}_{X|U}\widetilde{Q}_{Y|U}|\widetilde{Q}_{U})\big)\nonumber \\
 & \qquad+D(\widetilde{Q}_{XY|U}\|\pi_{XY}|\widetilde{Q}_{U}).\label{eq:-2}
\end{align}
Observe that $R_{\mathrm{sh}}$ is finite due to \eqref{eq:-8}. Hence
it holds that
\begin{align}
D(\widetilde{Q}_{XY}\|\pi_{XY}) & =0,\\
D(\widetilde{Q}_{XY|U}\|\widetilde{Q}_{X|U}\widetilde{Q}_{Y|U}|\widetilde{Q}_{U}) & =0.
\end{align}
That is,
\begin{align}
\widetilde{Q}_{XY} & =\pi_{XY},\label{eq:-6}\\
\widetilde{Q}_{XY|U} & =\widetilde{Q}_{X|U}\widetilde{Q}_{Y|U}.\label{eq:-7}
\end{align}
Therefore, under \eqref{eq:-6} and \eqref{eq:-7}, we have
\begin{align}
\eqref{eq:-2} & \geq D(\widetilde{Q}_{XY|U}\|\pi_{XY}|\widetilde{Q}_{U})\\
 & =I(\widetilde{Q}_{XY|U},\widetilde{Q}_{XY})\\
 & \geq R^{*}.\label{eq:-1}
\end{align}
Combining \eqref{eq:-8}, \eqref{eq:-2} and \eqref{eq:-1} yields
us
\begin{equation}
R_{\mathrm{sh}}=R^{*}=\lim_{k\to\infty}\frac{1}{\alpha_{m_{k}}}R^{(\alpha_{m_{k}})}.
\end{equation}
Therefore, there exists some sequence $\{c(\alpha_{m_{k}})\}_{k=1}^{\infty}\subset\bbR$
(e.g., the sequence $\{R^{*}-\frac{1}{\alpha_{m_{k}}}R^{(\alpha_{m_{k}})}\}_{k=1}^{\infty}\subset\bbR$)
such that $\lim_{k\to\infty}c(\alpha_{m_{k}})=0$ and
\begin{equation}
R^{*}-c(\alpha_{m_{k}})\leq\frac{1}{\alpha_{m_{k}}}R^{(\alpha_{m_{k}})}\leq R^{*}.\label{eq:-20}
\end{equation}
This concludes the proof.
\end{IEEEproof}
We also have the following crucial lemma.
\begin{lem}
\label{lem:Omega} Let $\alpha\in(0,1]$ and $Q_{XYU}\in\calQ$. Then
we have
\begin{equation}
\lim_{\theta\downarrow0}\frac{1}{\theta}\Omega^{(\alpha,\theta)}=R^{(\alpha)},\label{eq:-56}
\end{equation}
or equivalently,
\begin{equation}
\frac{1}{\theta}\Omega^{(\alpha,\theta)}=R^{(\alpha)}+\epsilon^{(\alpha,\theta)},\label{eqn:theta_to_zero}
\end{equation}
where $\Omega^{(\alpha,\theta)}$ and $R^{(\alpha)}$ were defined
in \eqref{eq:Omega} and \eqref{eq:Ra} respectively, and $\epsilon^{(\alpha,\theta)}$
is a term that vanishes as $\theta\downarrow0$, the rate being dependent
on $\alpha$.
\end{lem}
\begin{IEEEproof}[Proof of Lemma \ref{lem:Omega}]
To show this lemma, we first need to show that
\begin{equation}
\widehat{R}^{(\alpha,\theta)}(Q_{XYU}):=\begin{cases}
\frac{1}{\theta}\Omega^{(\alpha,\theta)}(Q_{XYU}), & \theta>0\\
R^{(\alpha)}(Q_{XYU}), & \theta=0
\end{cases}
\end{equation}
is continuous in $(\theta,Q_{XYU})\in[0,\frac{1}{1+\bar{\alpha}})\times\calQ$.
It is easy to observe that
\begin{align}
 & \Omega^{(\alpha,\theta)}(Q_{XYU})\nonumber \\
 & =-\log\mathbb{E}_{Q_{XYU}}\Big[\exp\big(-\theta\omega_{Q_{XYU}}^{(\alpha)}(X,Y|U)\big)\Big]\\
 & =-\log\sum_{x,y,u}Q_{XYU}^{1-\theta(1+\bar{\alpha})}(x,y,u)\left(Q_{U}(u)\pi_{XY}(x,y)\right)^{\theta}\nonumber \\
 & \qquad\times\left(Q_{U|XY}(u|x,y)Q_{X|U}(x|u)Q_{Y|U}(y|u)\right)^{\theta\bar{\alpha}}
\end{align}
is jointly continuous in $(\theta,Q_{XYU})\in[0,\frac{1}{1+\bar{\alpha}})\times\calQ$,
hence $\widehat{R}^{(\alpha,\theta)}(Q_{XYU})$ is jointly continuous
on $(0,\frac{1}{1+\bar{\alpha}})\times\calQ$. Therefore, to show
the continuity of $\widehat{R}^{(\alpha,\theta)}(Q_{XYU})$ in $(\theta,Q_{XYU})\in[0,\frac{1}{1+\bar{\alpha}})\times\calQ$,
it suffices to show it is continuous at any point in $\{0\}\times\calQ$,
i.e.,
\begin{equation}
\lim_{(\theta,Q_{XYU})\to(0,Q_{XYU}')}\frac{1}{\theta}\Omega^{(\alpha,\theta)}(Q_{XYU})=R^{(\alpha)}(Q_{XYU}')\label{eq:-53}
\end{equation}
for any $Q_{XYU}'\in\calQ$.

Let
\begin{align}
 & Q_{XYU}^{(\alpha,\theta)}(x,y,u)\nonumber \\
 &:= \frac{Q_{XYU}(x,y,u)\exp\Big(-\theta\omega_{Q_{XYU}}^{(\alpha)}(x,y|u)\Big)}{\sum_{x,y,u}Q_{XYU}(x,y,u)\exp\Big(-\theta\omega_{Q_{XYU}}^{(\alpha)}(x,y|u)\Big)}.\label{def:Qambl-1}
\end{align}
Invoking the definition of $\Omega^{(\alpha,\theta)}(Q_{XYU})$ in
\eqref{eq:OmegaQ}, we obtain
\begin{align}
\frac{\partial\Omega^{(\alpha,\theta)}(Q_{XYU})}{\partial\theta} & =\mathbb{E}_{Q_{XYU}^{(\alpha,\theta)}}\Big[\omega_{Q_{XYU}}^{(\alpha)}(X,Y|U)\Big],\label{eq:firstd}
\end{align}
and
\begin{align}
\frac{\partial^{2}\Omega^{(\alpha,\theta)}(Q_{XYU})}{\partial\theta^{2}} & =-\mathrm{Var}_{Q_{XYU}^{(\alpha,\theta)}}\Big[\omega_{Q_{XYU}}^{(\alpha)}(X,Y|U)\Big].\label{eq:secondd}
\end{align}
Hence for fixed $Q_{XYU}\in\calQ$, we have
\begin{align}
\left.\frac{\partial\Omega^{(\alpha,\theta)}(Q_{XYU})}{\partial\theta}\right|_{\theta=0}=R^{(\alpha)}(Q_{XYU}) & >0,\\
\frac{\partial^{2}\Omega^{(\alpha,\theta)}(Q_{XYU})}{\partial\theta^{2}} & \leq0,
\end{align}
which implies that
\begin{equation}
\theta\frac{\partial\Omega^{(\alpha,\theta)}(Q_{XYU})}{\partial\theta}\leq\Omega^{(\alpha,\theta)}(Q_{XYU})\leq\theta R^{(\alpha)}(Q_{XYU}).\label{eq:-50}
\end{equation}

Furthermore, observe that $R^{(\alpha)}(Q_{XYU})$ is continuous in
$Q_{XYU}\in\calQ$, hence
\begin{equation}
\lim_{Q_{XYU}\to Q_{XYU}'}R^{(\alpha)}(Q_{XYU})=R^{(\alpha)}(Q_{XYU}').\label{eq:-51}
\end{equation}
On the other hand, observe that $\frac{\partial\Omega^{(\alpha,\theta)}(Q_{XYU})}{\partial\theta}$
given in \eqref{eq:firstd} is continuous in $(\theta,Q_{XYU})\in[0,\frac{1}{1+\bar{\alpha}})\times\calQ$.
Hence
\begin{align}
 & \lim_{(\theta,Q_{XYU})\to(0,Q_{XYU}')}\frac{\partial\Omega^{(\alpha,\theta)}(Q_{XYU})}{\partial\theta}\nonumber \\
 &\qquad =\sum_{x,y,u}Q_{XYU}'(x,y,u)\omega_{Q_{XYU}'}^{(\alpha)}(X,Y|U)\\
 &\qquad =R^{(\alpha)}(Q_{XYU}').\label{eq:-52}
\end{align}
Therefore, combining \eqref{eq:-50}, \eqref{eq:-51}, and \eqref{eq:-52},
we observe that the limit $\lim_{(\theta,Q_{XYU})\to(0,Q_{XYU}')}\frac{1}{\theta}\Omega^{(\alpha,\theta)}(Q_{XYU})$
exists, and moreover, $\lim_{(\theta,Q_{XYU})\to(0,Q_{XYU}')}\frac{1}{\theta}\Omega^{(\alpha,\theta)}(Q_{XYU})=R^{(\alpha)}(Q_{XYU}')$.
Hence, we obtain \eqref{eq:-53}. In other words, $\widehat{R}^{(\alpha,\theta)}(Q_{XYU})$
is jointly continuous in $(\theta,Q_{XYU})\in[0,\frac{1}{1+\bar{\alpha}})\times\calQ$.
In addition, observe that $\calQ$ is a compact set. By using the
following lemma we can assert that $\min_{Q_{XYU}\in\calQ}\widehat{R}^{(\alpha,\theta)}(Q_{XYU})$
is continuous in $\theta\in[0,\frac{1}{1+\bar{\alpha}})$.
\begin{lem}[Lemma~14 in~\cite{tan2011large}]
\label{lemma:continuity} Let $\mathcal{X}$ and $\mathcal{Y}$ be
two metric spaces and let $\calK\subset\mathcal{X}$ be a compact
set. Let $f:\mathcal{X}\times\mathcal{Y}\rightarrow\mathbb{R}$ be
a (jointly) continuous real-valued function. Then the function $g:\mathcal{Y}\rightarrow\mathbb{R}$,
defined as
\begin{equation}
g(y):=\min_{x\in\calK}\,f(x,y),\quad\forall\,y\in\mathcal{Y},\label{eqn:gy}
\end{equation}
is continuous on $\mathcal{Y}$.
\end{lem}
Considering the point $\theta=0$, we obtain
\begin{align}
 & \lim_{\theta\to0}\min_{Q_{XYU}\in\calQ}\widehat{R}^{(\alpha,\theta)}(Q_{XYU})\nonumber \\
 & =\min_{Q_{XYU}\in\calQ}\widehat{R}^{(\alpha,0)}(Q_{XYU})\\
 & =\min_{Q_{XYU}\in\calQ}R^{(\alpha)}(Q_{XYU})=R^{(\alpha)},\label{eq:-54}
\end{align}
where the first equality follows from Lemma \ref{lemma:continuity}
which essentially says that the limit and minimum operations can be
swapped. On the other hand, observe that
\begin{align}
 & \lim_{\theta\to0}\min_{Q_{XYU}\in\calQ}\widehat{R}^{(\alpha,\theta)}(Q_{XYU})\nonumber \\
 & =\lim_{\theta\to0}\min_{Q_{XYU}\in\calQ}\frac{1}{\theta}\Omega^{(\alpha,\theta)}(Q_{XYU})\\
 & =\lim_{\theta\to0}\frac{1}{\theta}\Omega^{(\alpha,\theta)}.\label{eq:-55}
\end{align}
Combining \eqref{eq:-54} and \eqref{eq:-55}, we obtain \eqref{eq:-56}
as desired.
\end{IEEEproof}

\subsection{Proof of Part (i) in Lemma \ref{propFOmega}}

Using Lemma \ref{lem:wyner}, we obtain that if $R<C_{\mathsf{Wyner}}(X;Y)$,
then
\begin{align}
R+\tau\leq R^{*}\label{usersheqran}
\end{align}
for some $\tau>0$. Further, invoking \eqref{eq:Ra-2} and \eqref{usersheqran},
we obtain that there exists $k_{0}$ such that for any $k\geq k_{0}$,
\begin{align}
R+\tau\leq\frac{1}{\alpha_{k}}R^{(\alpha_{k})}+c(\alpha_{k}),\label{uselbtilr}
\end{align}
and
\begin{align}
c(\alpha_{k})\leq\frac{\tau}{2}.\label{clossysmalltau}
\end{align}
Referring to \eqref{uselbtilr} and \eqref{clossysmalltau}, we obtain
that for any $k\geq k_{0}$,
\begin{align}
R+\frac{\tau}{2}\leq\frac{1}{\alpha_{k}}R^{(\alpha_{k})}.\label{readyforuse}
\end{align}
Therefore, invoking \eqref{def:lossyF}, we conclude that for any
$k\geq k_{0}$,
\begin{align}
F(R) & \geq\sup_{\theta\geq0}F^{(\alpha_{k},\theta)}(R)\\
 & \geq\sup_{\theta\in[0,\frac{1}{1+\bar{\alpha}_{k}})}F^{(\alpha_{k},\theta)}(R)\\
 & =\sup_{\theta\in[0,\frac{1}{1+\bar{\alpha}_{k}})}\frac{\Omega^{(\alpha_{k},\theta)}-\theta\alpha_{k}R}{1+(5-3\alpha_{k})\theta}\\
 & \geq\sup_{\theta\in[0,\frac{1}{1+\bar{\alpha}_{k}})}\frac{1}{1+5\theta}\Big\{\theta R^{(\alpha_{k})}+\theta\epsilon^{(\alpha_{k},\theta)}-\theta\alpha_{k}R\Big\}\label{usetaylordeftilr}\\
 & \geq\sup_{\theta\in[0,\frac{1}{1+\bar{\alpha}_{k}})}\frac{\theta}{1+5\theta}\Big\{\epsilon^{(\alpha_{k},\theta)}+\frac{\alpha_{k}\tau}{2}\Big\}\label{useready}\\
 & \geq\sup_{\theta\in[0,\widetilde{\theta}]}\frac{\alpha_{k}\tau\theta}{4(1+5\theta)}\label{eq:-17}\\
 & \geq\frac{\alpha_{k}\tau\widetilde{\theta}}{4(1+5\widetilde{\theta})},\label{basiccalcu}
\end{align}
where \eqref{usetaylordeftilr} follows from Lemma \ref{lem:Omega}
and the inequality $1+(5-3\alpha_{k})\theta\leq1+5\theta$, \eqref{useready}
follows from \eqref{readyforuse}, and \eqref{eq:-17} follows since
there exists a sufficiently small $\widetilde{\theta}\in(0,\frac{1}{1+\bar{\alpha}_{k}})$
such that $|\epsilon^{(\alpha_{k},\theta)}|\leq\frac{1}{4}{\alpha_{k}\tau}$
for all $\theta\leq\widetilde{\theta}$. Since the expression in \eqref{basiccalcu}
is positive, we have $F(R)>0$ as desired.

\subsection{Proof of Part (ii) in Lemma \ref{propFOmega}}

Because $\exp(\cdot)$ is convex, applying Jensen's inequality, we
obtain
\begin{align}
\Omega^{(\alpha,\theta)}(Q_{XYU}) & \leq\theta\mathbb{E}_{Q_{XYU}}\Big[\omega_{Q_{XYU}}^{(\alpha)}(X,Y|U)\Big]\label{readyuse2}\\
 & =\theta R^{(\alpha)}(Q_{XYU}).
\end{align}
Hence we have
\begin{align}
\Omega^{(\alpha,\theta)} & \leq\min_{Q_{XYU}\in\calQ}\theta R^{(\alpha)}(Q_{XYU})\\
 & =\theta R^{(\alpha)}.
\end{align}
Thus, recalling the definition of $F^{(\alpha,\theta)}(R)$ in \eqref{def:lossyFalphamubetalambda},
we obtain that
\begin{align}
F^{(\alpha,\theta)}(R) & =\frac{\Omega^{(\alpha,\theta)}-\theta\alpha R}{1+(5-3\alpha)\theta}\\*
 & \leq\frac{\theta\alpha(\frac{1}{\alpha}R^{(\alpha)}-R)}{1+(5-3\alpha)\theta}\\
 & \leq\frac{\theta\alpha(R_{\mathrm{sh}}-R)}{1+(5-3\alpha)\theta}\\
 & \leq0,\label{fnonpositive}
\end{align}
where \eqref{fnonpositive} follows from the assumption $R\geq C_{\mathsf{Wyner}}(X;Y)=R_{\mathrm{sh}}$.
On the other hand, note that
\begin{align}
\lim_{\theta\to0}F^{(\alpha,\theta)}=0.\label{fnonnegetive}
\end{align}
Hence, combining \eqref{fnonpositive} and \eqref{fnonnegetive},
we conclude that
\begin{align}
F & =\sup_{(\alpha,\theta)\in[0,1]\times[0,\infty)}F^{(\alpha,\theta)}(R)=0.
\end{align}

\subsection*{Acknowledgements}
The authors thank the reviewers and the editor for their suggestions to enhance the quality of the paper.
\bibliographystyle{unsrt}
\bibliography{ref}

\begin{IEEEbiographynophoto}{Lei Yu} received the B.E. and Ph.D. degrees, both in electronic engineering, from University of Science and Technology of China (USTC) in 2010 and 2015, respectively. From 2015 to 2017, he was a postdoctoral researcher at the Department of Electronic Engineering and Information Science (EEIS), USTC. Currently, he is a research fellow at the Department of Electrical and Computer Engineering, National University of Singapore. His research interests include information theory, probability theory, and security.
\end{IEEEbiographynophoto}
\begin{IEEEbiographynophoto}{Vincent Y.\ F.\ Tan} (S'07-M'11-SM'15) was born in Singapore in 1981. He is currently an Associate Professor in the Department of Electrical and Computer Engineering  and the Department of Mathematics at the National University of Singapore (NUS). He received the B.A.\ and M.Eng.\ degrees in Electrical and Information Sciences from Cambridge University in 2005 and the Ph.D.\ degree in Electrical Engineering and Computer Science (EECS) from the Massachusetts Institute of Technology in 2011.  His research interests include information theory and machine learning.

Dr.\ Tan received the MIT EECS Jin-Au Kong outstanding doctoral thesis prize in 2011, the NUS Young Investigator Award in 2014, the NUS Engineering Young Researcher Award in 2018, and the Singapore National Research Foundation (NRF) Fellowship (Class of 2018). He has authored a research monograph on {\em ``Asymptotic Estimates in Information Theory with Non-Vanishing Error Probabilities''} in the Foundations and Trends in Communications and Information Theory Series (NOW Publishers). He is currently an Editor of the IEEE Transactions on Communications and a Guest Editor for the IEEE Journal of Selected Topics in Signal Processing.
\end{IEEEbiographynophoto}

\end{document}